\documentclass[12pt,a4paper]{article}
\usepackage{amsmath,amssymb,graphicx,epsfig,cite,color}
\usepackage[left=2.7cm,right=2.7cm,top=2.5cm,bottom=2.5cm]{geometry}
\usepackage{subfigure}
\usepackage{multirow}
\usepackage[table]{xcolor}
\usepackage{booktabs}

\usepackage{xcolor,colortbl}

\usepackage{graphicx}
\usepackage{lscape}
\usepackage{appendix}
\usepackage{multirow}
\usepackage[colorlinks=true,linkcolor=purple,citecolor=purple]{hyperref}%
\allowdisplaybreaks[1]

\begin{document}

\def\ds{\displaystyle}
\def\beq{\begin{equation}}
\def\eeq{\end{equation}}
\def\bea{\begin{eqnarray}}
\def\eea{\end{eqnarray}}
\def\beeq{\begin{eqnarray}}
\def\eeeq{\end{eqnarray}}

\def\rar{\rightarrow} 
\def\nnb{\nonumber}

\def\ds{\displaystyle}
\def\beq{\begin{equation}}
\def\eeq{\end{equation}}
\def\bea{\begin{eqnarray}}
\def\eea{\end{eqnarray}}
\def\beeq{\begin{eqnarray}}
\def\eeeq{\end{eqnarray}}
\def\ve{\vert}
\def\vel{\left|}
\def\ver{\right|}
\def\nnb{\nonumber}
\def\ga{\left(}
\def\dr{\right)}
\def\aga{\left\{}
\def\adr{\right\}}
\def\lla{\left<}
\def\rra{\right>}
\def\rar{\rightarrow}
\def\lrar{\leftrightarrow}
\def\nnb{\nonumber}
\def\la{\langle}
\def\ra{\rangle}
\def\ba{\begin{array}}
\def\ea{\end{array}}
\def\tr{\mbox{Tr}}
\def\ssp{{\Sigma^{*+}}}
\def\sso{{\Sigma^{*0}}}
\def\ssm{{\Sigma^{*-}}}
\def\xis0{{\Xi^*}}
\def\xism{{\Xi^{*-}}}
\def\qs{\la \bar s s \ra}
\def\qu{\la \bar u u \ra}
\def\qd{\la \bar d d \ra}
\def\qq{\la \bar q q \ra}
\def\gGgG{\la g^2 G^2 \ra}
\def\q{\gamma_5 \not\!q}
\def\x{\gamma_5 \not\!x}
\def\g5{\gamma_5}
\def\sb{S_Q^{cf}}
\def\sd{S_d^{be}}
\def\su{S_u^{ad}}
\def\sbp{{S}_Q^{'cf}}
\def\sdp{{S}_d^{'be}}
\def\sup{{S}_u^{'ad}}
\def\ssp{{S}_s^{'??}}

\def\sig{\sigma_{\mu \nu} \gamma_5 p^\mu q^\nu}
\def\fo{f_0(\frac{s_0}{M^2})}
\def\ffi{f_1(\frac{s_0}{M^2})}
\def\fii{f_2(\frac{s_0}{M^2})}
\def\O{{\cal O}}
\def\sl{{\Sigma^0 \Lambda}}
\def\es{\!\!\! &=& \!\!\!}
\def\ap{\!\!\! &\approx& \!\!\!}
\def\ar{&+& \!\!\!}
\def\ek{&-& \!\!\!}
\def\kek{\!\!\!&-& \!\!\!}
\def\cp{&\times& \!\!\!}
\def\se{\!\!\! &\simeq& \!\!\!}
\def\eqv{&\equiv& \!\!\!}
\def\kpm{&\pm& \!\!\!}
\def\kmp{&\mp& \!\!\!}
\def\mcdot{\!\cdot\!}
\def\erar{&\rightarrow&}



\renewcommand{\textfraction}{0.2}    
\renewcommand{\topfraction}{0.8}   

\renewcommand{\bottomfraction}{0.4}   
\renewcommand{\floatpagefraction}{0.8}
\newcommand\mysection{\setcounter{equation}{0}\section}

\def\baeq{\begin{appeq}}     \def\eaeq{\end{appeq}}  
\def\baeeq{\begin{appeeq}}   \def\eaeeq{\end{appeeq}}
\newenvironment{appeq}{\beq}{\eeq}   
\newenvironment{appeeq}{\beeq}{\eeeq}
\def\bAPP#1#2{
 \markright{APPENDIX #1}
 \addcontentsline{toc}{section}{Appendix #1: #2}
 \medskip
 \medskip
 \begin{center}      {\bf\LARGE Appendix #1 :}{\quad\Large\bf #2}
\end{center}
 \renewcommand{\thesection}{#1.\arabic{section}}
\setcounter{equation}{0}
        \renewcommand{\thehran}{#1.\arabic{hran}}
\renewenvironment{appeq}
  {  \renewcommand{\theequation}{#1.\arabic{equation}}
     \beq
  }{\eeq}
\renewenvironment{appeeq}
  {  \renewcommand{\theequation}{#1.\arabic{equation}}
     \beeq
  }{\eeeq}
\nopagebreak \noindent}

\def\eAPP{\renewcommand{\thehran}{\thesection.\arabic{hran}}}

\renewcommand{\theequation}{\arabic{equation}}
\newcounter{hran}
\renewcommand{\thehran}{\thesection.\arabic{hran}}

\def\bmini{\setcounter{hran}{\value{equation}}
\refstepcounter{hran}\setcounter{equation}{0}
\renewcommand{\theequation}{\thehran\alph{equation}}\begin{eqnarray}}
\def\bminiG#1{\setcounter{hran}{\value{equation}}
\refstepcounter{hran}\setcounter{equation}{-1}
\renewcommand{\theequation}{\thehran\alph{equation}}
\refstepcounter{equation}\label{#1}\begin{eqnarray}}


\newskip\humongous \humongous=0pt plus 1000pt minus 1000pt
\def\caja{\mathsurround=0pt}
 

\title{
         {\Large
                 {\bf
Strong interaction  of doubly heavy spin-3/2 baryons with light  vector mesons  
                 }
         }
      }

\author{\vspace{1cm}\\
	{\small
		K. Azizi$^{1,2,4}$,
		A. R. Olamaei$^{3,4}$,  
		S.~Rostami$^{1}$} \\
	{\small $^1$ Department of Physics, University of Tehran, North Karegar Ave. Tehran 14395-547, Iran}\\
	{\small $^2$ Department of Physics, Do\v{g}u\c{s} University, Acibadem-Kadik\"{o}y, 34722
		Istanbul, Turkey}\\
	{\small$^3$  Department of Physics, Jahrom University, Jahrom, P.~ O.~ Box 74137-66171, Jahrom, Iran}\\
	{\small $^4$ School of Particles and Accelerators, Institute for Research in Fundamental 
		Sciences (IPM),}\\
	{\small  P. O. Box 19395-5531, Tehran, Iran}\\
	} 

\date{}

\begin{titlepage}
\maketitle
\thispagestyle{empty}

\begin{abstract}
We calculate the strong coupling constants among  the
doubly heavy spin-$ \frac{3}{2} $ baryons $\Xi^*_{QQ}$  and  $\Omega^*_{QQ}$, with $ Q $ and $ Q' $ being $  c$ or $ b $ quark,  with light  vector meson by means of the light-cone QCD sum rules. The matrix elements defining these vertices are described by four coupling constants $ g_1$, $ g_2$, $ g_3$, and  $ g_4 $. The unwanted pollution coming from the doubly heavy spin-$ \frac{1}{2} $ baryons are removed by a special ordering of Dirac matrices and selection of appropriate Lorentz structures. The strong coupling constants are basic parameters that carry information on the nature of the strong interaction among hadronic multiplets. Investigation of these parameters may help physicists in the construction of the strong potentials among the doubly heavy baryons and light vector mesons. The values obtained for the strong coupling constants may also help experimental groups in analyses of the data produced at various hadron colliders. 

\end{abstract}

\end{titlepage}
\section{Introduction}
Studies on the spectroscopic and various decay parameters as well as  internal structures of doubly heavy baryons constitute one of the directions of active research in particle physics.
In the last two decades, many aspects of these baryons which consist of two heavy and one light quarks,
 have been studied both theoretically and experimentally \cite{Wang:2017mqp,Wang:2017azm,
Gutsche:2017hux,Li:2017pxa,Xiao:2017udy,Sharma:2017txj,Ma:2017nik,Hu:2017dzi,Shi:2017dto,Yao:2018ifh,Zhao:2018mrg,Liu:2018euh,Xing:2018lre,Dhir:2018twm,Berezhnoy:2018bde,Jiang:2018oak,Zhang:2018llc,Gutsche:2018msz,Shi:2019fph,Hu:2019bqj,Brodsky:2011zs,Mattson:2002vu,Ocherashvili:2004hi,Ratti:2003ez,Aubert:2006qw,Aaij:2013voa,Kato:2013ynr,Aaij:2017ueg,Aliev:2012ru,Aliev:2012iv,Azizi:2014jxa,Aliev:2012nn,Ozdem:2019zis}.
 From the experimental point of view, 
  a new era began in the studies of double
charm baryons when the SELEX Collaboration at Fermilab reported the observation of 
$ \Xi _{cc}^+ $ baryon  in the decay modes $ \Lambda_c^+ K^- \pi^+ $
and $ p D^+ K^- $ with  mass $ M_{\Xi _{cc}^+ }  = (3518.7\pm1.7)$ MeV \cite{Mattson:2002vu,Ocherashvili:2004hi}.
However, in the years that followed, other collaborations like FOCUS \cite{Ratti:2003ez}, BaBar \cite{Aubert:2006qw}, LHCb \cite{Aaij:2013voa}, and Belle \cite{Kato:2013ynr} failed to observe such a state with the properties that the  SELEX observed. 
Finally, the LHCb Collaboration discovered the double charm state $ \Xi^{++}_{cc} $  in 2017 \cite{Aaij:2017ueg}
and confirmed via measuring  the decay channel $\Xi^{++}_{cc}\rightarrow \Xi^+_c \pi^+$ \cite{Aaij:2018gfl}.
The extracted mass of the $ \Xi^{++}_{cc} $ state was given as
$3621.24 \pm 0.65 (\text{stat.})\pm 0.31 (\text{syst.})~\text{MeV}/c^2 $.
Brodsky et al. \cite{Brodsky:2017ntu} found that despite the difference between 
the LHCb and SELEX values for the mass  (about $100~ \text{MeV}/c^2 $), 
 both can  be correct with application of supersymmetric algebra to hadron spectroscopy, together with the intrinsic heavy-quark QCD mechanism for the hadroproduction of heavy hadrons at large $ x_F$.
 
 Recently, 
 with a data sample corresponding to an integrated luminosity  of 5.4~\ensuremath{\mbox{fb}^{-1}}, 
the $\Xi^0_{bc}\rightarrow D^0 p K^- $ decay has been searched for by the LHCb and no evidence for a signal is found \cite{Aaij:2020vid}.
Of course, the story will not end here and experiments will continue.
Future searches at LHCb with improved trigger conditions, and larger data samples will further improve the sensitivity of
 doubly heavy baryon  signals.
 
 Given that  more doubly heavy baryons  are expected to be discovered experimentally in the future,  theoretical studies on different properties of them are needed to provide many
inputs to experiments. We hope that not only we will  be able to identify the ground state doubly heavy baryons predicted by the   quark model \cite{GELLMANN1964214,Zweig:1981pd,Zweig:1964jf}, but also will find more information on the resonances beyond the ground state particles.
Heretofore, lattice QCD \cite{Brown:2014ena}, Quark Spin Symmetry \cite{Hernandez:2007qv,Flynn:2011gf}, QCD Sum Rules \cite{ Aliev:2012ru,Aliev:2012iv,Azizi:2014jxa,Azizi:2018duk,Olamaei:2020bvw}  are just some of the attempts that have been made to predict masses,  residues,  decay
modes, lifetimes, strong coupling constants, and other properties of doubly heavy baryons.

In the present work, we aim to study 
 the strong coupling constants among the doubly heavy spin-3/2 baryons $ \Xi^*_{QQ'}, \Omega^*_{QQ'} $ and light vector mesons $ \rho, \omega, \phi,$ and $ K^{*} $.
The formation of hadrons takes place at low energy
scale, which belongs to the nonperturbative regime of QCD. The strong coupling constants among the hadrons are basic quantities that appear when the hadrons meet each other at colliders. For the calculations of these fundamental parameters, 
one can not use the fundamental Lagrangian of QCD and needs some nonperturbative approaches.
One of the most powerful approaches in this respect is the QCD sum rules method \cite{Shifman:1978bx}.
 In the present work, we investigate the strong coupling constants of the doubly heavy spin-3/2 baryons with light vector mesons within the light cone QCD sum rule's (LCSR) framework.
 The LCSR approach adopts the operator product expansion (OPE) near the light cone  $ x^2\approx 0 $ instead of the short distance $ x\approx 0 $, whose matrix elements are parametrized by the hadronic DA’s which are classified according to their twists \cite{Balitsky:1989ry,Khodjamirian:1997ub,Braun:1997kw}. Note that the strong coupling constants among the spin-1/2 baryons and light mesons are investigated in \cite{Olamaei:2020bvw,Rostami:2020euc,Alrebdi:2020rev,Aliev:2020aon}.

The plan for this article is as follows. In next section, we
   review the formalism of driving the sum rules for the strong coupling constants in the LCSR using both the physical and QCD representations of the correlation function (CF).  In Section \ref{NA}, the numerical analysis
 and numerical results for the strong coupling constants are presented. Section \ref{SC} is devoted to a brief summary and conclusion.

\section{LCSR for the strong coupling constants }\label{LH}
In this section, we aim to  construct the LCSR for the strong coupling constants of the doubly heavy baryons $\Xi^*_{QQ}$  and  $\Omega^*_{QQ}$
with light vector mesons. To this end, the main inputs are the interpolating currents of the doubly heavy baryons in terms of the quark fields inside them. In a compact form, the interpolating current for spin-3/2 doubly heavy baryons  can be written as 
\begin{eqnarray}\label{eta}
\eta_{\mu}(x) &=&\frac{\epsilon_{abc}}{\sqrt{3}} \Bigg\{[q^{aT}(x)C \gamma_{\mu}Q^b(x)]Q'^c(x)+
[q^{aT}(x)C  \gamma_{\mu} Q'^b(x)]Q^c(x)\\ \nonumber
&+&[Q^{aT}(x)C \gamma_{\mu} Q'^b(x)]q^c(x) \Bigg\},
\end{eqnarray}
where $ \epsilon $ stands for the anti-symmetric Levi-Civita tensor with $ a, b $ and $ c $ being the color indexes.
$C$ and $T$ are the charge conjugation and transposition operators, respectively. 
The quark contents for the doubly heavy spin-3/2 baryons family are shown in Table \ref{tab:baryon3/2}.

\begin{table}
	\setlength{\tabcolsep}{1.45em} 
	\centering
	\begin{tabular}{cccc}
		\hline
		\hline
		\textcolor{purple}{Baryon}    &\textcolor{purple}{ $ q $ }& \textcolor{purple}{$ Q $ }& \textcolor{purple}{$ Q^\prime $ }\\
		\hline
		\hline
		\\
		$  \Xi^*_{QQ^\prime }$     & $ u $ or $ d $    & $b  $   or $ c $ & $b  $   or $ c $    \\
		&         &  &     \\
		$ \Omega^*_{QQ^\prime}$   & $ s $   & $b  $   or $ c $ & $b  $   or $ c $       \\
		\\
		\hline
	\end{tabular}
	\caption{The quark content of the doubly heavy spin-3/2 baryons.} \label{tab:baryon3/2} 
\end{table}

To derive LCSR for  the strong vertices of 
$  \Xi^*_{QQ^\prime }$ or $ \Omega^*_{QQ^\prime}$ baryons with the light vector mesons $ \rho,  K^{*}, \omega $ or $\phi$, the starting point is to consider the following CF:
\begin{eqnarray}\label{CF} 
\Pi_{\mu \nu}= i \int d^4x e^{ip.x} \left< V(q) \vert {\cal T} \left\{
\eta_{\mu} (x) \bar{\eta}_{\nu} (0) \right\} \vert 0 \right>~,
\end{eqnarray}
where the time-ordered product of two  currents located between an on-shell vector meson state $  V(q)  $ with four momentum $ q $ and the hadronic vacuum. $p$ is the external four-momentum of the outgoing doubly heavy baryon,
and $\eta (x)  $ and $ \bar{\eta} (0) $ are interpolating currents for doubly heavy spin-3/2 baryons. This  correlation function is calculated both at hadronic (low energy) and QCD (high energy) levels:
\begin{itemize}
	\item[$ \bullet $] At the hadron level (in the time-like region), the CF is calculated by inserting  complete sets of  
	baryonic states with the same quantum numbers as the interpolating 
	currents  and isolating the ground state contribution. It is called the phenomenological or physical representation of the CF.

	\item[$ \bullet $] At the QCD level (in the space-like region), the CF is calculated in terms of QCD degrees of freedom
	by using the operator product expansion (OPE).
	It is known as the QCD or theoretical representation of the CF. 
	
\end{itemize}
By matching these two representations via a dispersion relation, we can get the desired sum rules.
By applying the Borel transformation and continuum subtraction procedures 
one can remove the  divergences coming from the dispersion integrals as well as suppress the contributions of the  higher states and continuum.

Thus, in the hadronic representation,    complete sets of baryonic states  are placed  in the CF. By performing the Fourier integration over $x$ and isolating the  ground state contribution we get

\begin{eqnarray}\label{hadron-level} 
\Pi^{\text{Phys.}}_{\mu\nu}(p,q)=\frac{\langle 0\vert \eta_\mu \vert B^*_2(p,r)\rangle  \langle B^*_2(p,r)V(q)\vert B^*_1(p+q,s)\rangle\langle B^*_1(p+q,s) \vert \bar{\eta_\mu}\vert 0\rangle}{(p^2-m_2^2)[(p+q)^2-m_1^2]} +\cdots~,
\end{eqnarray}
where dots correspond to the contribution of the  higher states and continuum. $B^*_1(p+q,s)$ and $B^*_2(p,r)$ are the initial and final doubly heavy  baryons with spins $s$ and $r$, respectively.
To proceed, we introduce
\begin{eqnarray}\label{me1} 
\langle 0\vert \eta_\mu\vert B^*_i(p,s)\rangle &=&\lambda_{B^*_i}u_\mu(p,s),
\end{eqnarray}
where $\lambda_{B^*_i}$ acts as the residue and $u_\mu(p,s)$ are the  Rarita--Schwinger spinor for the baryon $B^*_i$ with momentum $p$ and spin $s$. Using the Lorentz and parity symmetries one can define the matrix element $\lla B_2^*(p,r) V(q)
\vel \right. B_1^*(p+q,s) \rra$ as (see for instance Refs. \cite{Aliev:2011uf,Pascalutsa:2006up}):
\begin{eqnarray}
\label{eBQsBQV05}
\lla B_2^*(p,r) V(q) \vel \right. B_1^*(p+q,s) \rra \es \bar{u}_\alpha (p,r)
\Bigg\{ g^{\alpha\beta} \Bigg[ \rlap/\varepsilon g_1 + 2 (p.\varepsilon) {g_2
	\over m_1+m_2} \Bigg] \nnb \\
\ar {q^\alpha q^\beta \over (m_1+m_2)^2} \Bigg[\rlap/\varepsilon g_3 + 2 (p.\varepsilon) {g_4\over m_1+m_2} \Bigg] \Bigg\} u_\beta (p+q,s),
\end{eqnarray}
in terms of four strong coupling  form factors $g_1, g_2, g_3$ and $g_4$. Here  $\varepsilon_\mu$ is the polarization four-vector of the vector meson $V(q)$. 

Inserting Eqs. \ref{eBQsBQV05} and \ref{me1} into Eq. \ref{hadron-level}, since the baryons are considered as un-polarized, the next stage is to sum over the spins of the incoming and outgoing baryons which can be done using the following completeness relation:
\begin{eqnarray}
\label{eBQsBQV06}
\sum_s u_\mu (p,s) \bar{u}_\nu (p,s) = ( {\rlap/p +m } )\Bigg(
g_{\mu\nu} - {1\over 3} \gamma_\mu \gamma_\nu + {2 p_\mu p_\nu \over 3
	m^2} + {p_\mu \gamma_\nu - p_\nu \gamma_\mu \over 3 m} \Bigg)~.
\end{eqnarray} 
Now, it is necessary to note two important points. First, one should note that the appearing structures are not all independent. The second point is that there are contributions from spin-1/2 states due to the coupling of the interpolating current $\eta_{\mu}$ to them via
\begin{eqnarray}
\label{eBQsBQV07}
\lla 0 \vel \eta_{\mu}\ver 1/2(p) \rra \es A \ga \gamma_\mu - {4\over m}
p_\mu \dr u(p)~.
\end{eqnarray} 
Therefore one can find out from the above equation that the structures proportional to $\gamma_{\mu}$ from the left or $\gamma_{\nu}$ from the right, and also the ones proportional to $p_{\mu}$ and $(p+q)_{\nu}$ receive contributions from the unwanted spin-1/2 states which have to be removed. Both requirements can be fulfilled by reordering  the  Dirac matrices. In this work we choose  $\gamma_\mu \rlap/\varepsilon \rlap/q \rlap/p \gamma_\nu$ as the suitable ordering and remove the spin-1/2 pollution using the above prescription. 

Putting all the above things together, the final form of the physical side of the CF is
\begin{eqnarray}
\label{eBQsBQV08}
\Pi^{\text{Phys.}}_{\mu\nu}(p,q) \es { \lambda_{B_{1}^*} \lambda_{B_{2}^*} \over
	[(p+q)^2-m_1^2)] (p^2 -  m_2^2)} \Bigg\{
2 (\varepsilon.p) g_{\mu\nu} \rlap/q \Bigg[ g_1 + 
g_2 {m_2\over m_1+m_2}\Bigg] \nnb \\
\ek 2 (\varepsilon.p) g_{\mu\nu} \rlap/q \rlap/p {g_2\over
	m_1+m_2} + q_\mu q_\nu \rlap/\varepsilon \rlap/q \rlap/p 
{g_3 \over (m_1+m_2)^2} -
2 (\varepsilon.p) q_\mu q_\nu \rlap/q \rlap/p {g_4 \over
	(m_1+m_2)^3} \nnb \\
\ar \mbox{\rm other structures} +...\Bigg\}~.
\end{eqnarray}
To acquire the sum rules for the  strong coupling constants $g_i$, we choose the coefficients of the structures  $(\varepsilon.p)
g_{\mu\nu} \rlap/q$, $(\varepsilon.p) g_{\mu\nu} \rlap/q \rlap/p$,
$q_\mu q_\nu \rlap/\varepsilon \rlap/q \rlap/p$ and $(\varepsilon.p) q_\mu
q_\nu \rlap/q \rlap/p$ from both the hadronic and QCD sides.

The theoretical (QCD) side of the CF is calculated in deep Euclidean region, where $-p^2 \rar \infty$, $-(p+q)^2 \rar \infty$. To separate the perturbative and non-perturbative contributions we employ the OPE. The non-perturbative inputs in LCSR are distribution amplitudes (DAs) of the corresponding light vector mesons which can be found in \cite{Ball:1998sk,Ball:1998ff,Ball:2007zt}.

To proceed in QCD side, we insert the interpolating current \ref{eta} to the CF \ref{CF} and use the Wick theorem to find
\begin{eqnarray}\label{CF1}
\Big(\Pi^{\text{QCD}}_{\mu \nu}\Big)_{\rho\tau}(p,q) &=&  \frac{i}{3}\epsilon_{abc}\epsilon_{a'b'c'} \int d^4 x e^{i p.x} \langle V(q) \vert \bar{q}^{c^\prime}_{\alpha}(0)q^{c}_{\beta}(x)\vert 0\rangle \Bigg\{\delta_{ \alpha\rho} \delta_{\beta \tau }\text{Tr} \Big(  \tilde{S}^{aa^{\prime}}_{Q}(x) \gamma_\mu S^{bb^{\prime}}_{Q^{\prime}}(x) \gamma_\nu  \Big) \nonumber\\
&+&
\delta_{\alpha \rho}\Big(  \gamma_\nu \tilde{S}^{aa^{\prime}}_{Q}(x) \gamma_\mu S^{bb^{\prime}}_{Q^{\prime}}(x) \Big)_{\beta \tau } 
+
\delta_{\alpha \rho} \Big( \gamma_\nu   \tilde{S}^{bb^{\prime}}_{Q^{\prime}}(x) \gamma_\mu S^{aa^{\prime}}_{Q}(x) \Big)_{\beta \tau } 
\nonumber\\
&+&
\delta_{ \beta\tau} \Big(  S^{bb^{\prime}}_{Q^{\prime}}(x)  \gamma_\nu \tilde{S}^{aa^{\prime}}_{Q}(x)  \gamma_\mu \Big)_{ \rho \alpha } 
+
\Big(  \gamma_\nu \tilde{S}^{aa^{\prime}}_{Q}(x)  \gamma_\mu   \Big)_{ \beta \alpha} \Big(S^{bb^{\prime}}_{Q^{\prime}}(x)\Big)_{\rho \tau}
\nonumber\\
&-&
\Big( C  \gamma_\mu S^{aa^{\prime}}_{Q}(x)  \Big)_{\alpha \tau} \Big(   S^{bb^{\prime}}_{Q^{\prime}}(x) \gamma_\nu C \Big)_{\rho \beta} -
\delta_{\beta \tau}\Big( S^{aa^{\prime}}_{Q}(x)  \gamma_\nu   \tilde{S}^{bb^{\prime}}_{Q^{\prime}}(x) \gamma_\mu  \Big)_{ \rho \alpha}
\nonumber\\
&-&
\Big( C  \gamma_\mu S^{bb^{\prime}}_{Q^{\prime}}(x)  \Big)_{\alpha \tau}  \Big(   S^{aa^{\prime}}_{Q}(x) \gamma_\nu C \Big)_{\rho \beta} 
+
\Big(\gamma_\nu  \tilde{S}^{bb^{\prime}}_{Q^{\prime}}(x)  \gamma_\mu \Big)_{\beta \alpha } \Big(S^{aa^{\prime}}_{Q}(x)\Big)_{\rho \tau}\Bigg\},
\end{eqnarray}
in terms of the heavy quark propagators, $S^{aa^{\prime}}_{Q{^{(\prime)}}}(x)$. Here, $\tilde{S} = C S^T C$ and the main non-perturbative contributions are packed in the non-local matrix elements $\langle V(q) \vert \bar{q}^{c^\prime}_{\alpha}(x)q^{c}_{\beta}(0)\vert 0\rangle$, which can be expressed in terms of DAs of the light vector meson $V(q)$. 

The heavy quark propagator can be written in terms of the free and interacting parts as
\begin{eqnarray}\label{HQP}
S_Q^{aa^{\prime}}(x) &=& {m_Q^2 \over 4 \pi^2} {K_1(m_Q\sqrt{-x^2}) \over \sqrt{-x^2}}\delta^{aa^{\prime}} -
i {m_Q^2 \rlap/{x} \over 4 \pi^2 x^2} K_2(m_Q\sqrt{-x^2})\delta^{aa^{\prime}}\nnb \\& -&
ig_s \int {d^4k \over (2\pi)^4} e^{-ikx} \int_0^1
du \Bigg[ {\rlap/k+m_Q \over 2 (m_Q^2-k^2)^2} \sigma^{\lambda\theta} G_{\lambda\theta}^{aa^{\prime}} (ux)
\nnb \\
&+&
{u \over m_Q^2-k^2} x^\lambda  \gamma^\theta G_{\lambda\theta}^{aa^{\prime}}(ux) \Bigg]+\cdots,
\end{eqnarray}
where the first two terms represent the propagation of free quark, in which $K_1$ and $K_2$ are the modified Bessel functions of the second kind. The rest, which  $\sim G_{\lambda\theta}^{aa^{\prime}}$, represent the interacting part.  Here $ G_{\lambda\theta}^{aa^{\prime}}$ is the gluon strength tensor with the shorthand notation 
\begin{eqnarray}
G^{aa^{\prime}}_{\lambda\theta }\equiv G^{A}_{\lambda\theta}t^{aa^{\prime}}_{A},
\end{eqnarray}
where $\lambda$ and $\theta$ are Minkowski indices, $t_{A} = \lambda_{A}/2$ and $\lambda^{aa^{\prime}}_A$ is the $aa^{\prime}$th array of the Gell-Mann tensor $\lambda_A$ in which $A = 1,2, ... , 8$.

As the heavy quark propagator \ref{HQP} consists of both free and gluonic terms, it leads to several contributions in the CF as said. The leading order contribution is obtained by  replacing both the heavy quark propagators with their free parts
\begin{eqnarray}\label{HQPpert}
S_Q^{aa^{\prime}(\text{pert.})}(x) &=& {m_Q^2 \over 4 \pi^2} {K_1(m_Q\sqrt{-x^2}) \over \sqrt{-x^2}} \delta^{aa^{\prime}}-
i {m_Q^2 \rlap/{x} \over 4 \pi^2 x^2} K_2(m_Q\sqrt{-x^2})\delta^{aa^{\prime}}.
\end{eqnarray}  
It represents a bare loop with no gluon exchange and the non-local matrix elements in this case can be expressed in terms of two-particle DAs of the light vector meson $V(q)$. 
Inserting the free part of one heavy quark propagator and the gluonic part of the other, which is
\begin{eqnarray}\label{HQPnp}
S^{aa^{\prime}(\text{non-p.})}_{Q}(x)&=&-
ig_s \int {d^4k \over (2\pi)^4} e^{-ikx} \int_0^1
du G^{aa^{\prime}}_{\mu\nu}(ux) \Delta^{\mu\nu}_{Q}(x),
\end{eqnarray}
where $ \Delta^{\mu\nu}_{Q}(x)$ is
\begin{eqnarray}\label{HQPgamma}
\Delta^{\mu\nu}_{Q}(x)&=& \dfrac{1}{2 (m_Q^2-k^2)^2}\Big[(\rlap/k+m_Q)\sigma^{\mu\nu} + 2u (m_Q^2-k^2)x^\mu \gamma^\nu\Big],
\end{eqnarray}
 leads to the contribution which is responsible for the exchange of one gluon between one of the heavy quarks and the light vector meson and can be calculated using three-particle DAs of the corresponding vector meson.   
The contribution coming from considering the gluonic parts of both the heavy quarks propagators representing the exchange of two gluons, leads to four-particle DAs which are not available yet and therefore we ignore it in the present study. However, we consider the two-gluon condensate contributions.

The non-local matrix elements can be calculated using the Fierz identities. For the quark part we have 
\begin{eqnarray}\label{eq:Fiertz1}
\bar q _\alpha ^{c'}(0) q _\beta ^c (x) \to - \frac{1}{12} (\Gamma _J) _{\beta\alpha} \delta ^{cc'} \bar q(0) \Gamma ^ J q(x) , 
\end{eqnarray}
and we use similar identity for the gluonic part.
Here, $\Gamma^J$ runs over all possible $\gamma$-matrices as
\begin{equation}\label{gammaexp}
\Gamma ^{J}=\mathbf{1,\ }\gamma _{5},\ \gamma _{\mu },\ i\gamma _{5}\gamma
_{\mu },\ \sigma _{\mu \nu }/\sqrt{2}.
\end{equation}
As a result of the above procedure, there appear the following two- and three-particle matrix elements
\begin{align}\label{eq:MatEl}
\langle V(q)|\bar q(0) \Gamma _J q(x)|0 \rangle ~~\mbox{and}~~  \langle V(q)|\bar q(0) \Gamma _J G _{\lambda \theta} (ux) q(x)|0\rangle ,
\end{align}
which can be expressed in terms of DAs of different twists. The expansion of the above matrix elements in terms of DAs as well as the explicit forms of the DAs for the vector mesons with different twists can be found in Refs. \cite{Ball:1998sk,Ball:1998ff,Ball:2007zt}.

Now, we briefly explain how different contributions to the corresponding form factors are calculated. The leading order contribution of the CF \ref{CF1}, which is obtained by replacing both the heavy quark propagators with the perturbative terms, is as follows 
\begin{eqnarray}\label{CFpert}
\Big(\Pi^{\text{QCD(0)}}_{\mu \nu}\Big)_{\rho\tau}(p,q) &=&  \frac{i}{6} \int d^4 x e^{i p.x} \langle V(q) \vert \bar{q}(0) \Gamma^J q(x)\vert 0\rangle \Bigg\{\Big(\Gamma_{J}\Big)_{\rho \tau}\text{Tr} \Big(  \tilde{S}^{(\text{pert.})}_{Q}(x) \gamma_\mu S^{(\text{pert.})}_{Q^{\prime}}(x) \gamma_\nu  \Big) \nonumber\\
&+&
\Big(\Gamma_J  \gamma_\nu \tilde{S}^{(\text{pert.})}_{Q}(x) \gamma_\mu S^{(\text{pert.})}_{Q^{\prime}}(x) \Big)_{\rho \tau } 
+
\Big( \Gamma_J \gamma_\nu   \tilde{S}^{(\text{pert.})}_{Q^{\prime}}(x) \gamma_\mu S^{(\text{pert.})}_{Q}(x)  \Big)_{\rho \tau } 
\nonumber\\
&+&
\Big(  S^{(\text{pert.})}_{Q^{\prime}}(x)  \gamma_\nu \tilde{S}^{(\text{pert.})}_{Q}(x)  \gamma_\mu \Gamma_J\Big)_{ \rho \tau } 
+
\text{Tr}  \Big( \Gamma_J \gamma_\nu \tilde{S}^{(\text{pert.})}_{Q}(x)  \gamma_\mu   \Big) \Big(S^{(\text{pert.})}_{Q^{\prime}}(x)\Big)_{\rho \tau}
\nonumber\\
&-&
\Big( S^{(\text{pert.})}_{Q^{\prime}}(x) \gamma_\nu  \tilde{\Gamma}_J  \gamma_\mu S^{(\text{pert.})}_{Q}(x)  \Big)_{\rho \tau}-
\Big( S^{(\text{pert.})}_{Q}(x)  \gamma_\nu   \tilde{S}^{(\text{pert.})}_{Q^{\prime}}(x) \gamma_\mu  \Gamma_J \Big)_{ \rho \tau}
\nonumber\\
&-&
\Big( S^{(\text{pert.})}_{Q}(x) \gamma_\nu \tilde{\Gamma}_J  \gamma_\mu S^{(\text{pert.})}_{Q^{\prime}}(x)  \Big)_{\rho \tau}  
+
\text{Tr} \Big(\Gamma_J \gamma_\nu  \tilde{S}^{(\text{pert.})}_{Q^{\prime}}(x)  \gamma_\mu \Big) \Big(S^{(\text{pert.})}_{Q}(x)\Big)_{\rho \tau}\Bigg\}, \nonumber\\
\end{eqnarray}
where the superscript $(0)$ indicates zero gluon exchange.
One can obtain the contribution of one gluon exchange (say between the heavy quark $Q$ and the vector meson) by replacing one heavy quark propagator ($Q'$) with its perturbative free terms and the other ($Q$) with its nonperturbative gluonic terms as 
\begin{eqnarray}\label{CFgluon}
\Big(\Pi^{\text{QCD(1)}}_{\mu \nu}\Big)_{\rho\tau}(p,q) &=&  \frac{-i g_s}{96} \int \frac{d^4 k}{(2\pi)^2} e^{-i k.x} \int_0^1 du  \langle V(q) \vert \bar{q}(0)\Gamma^J G_{\lambda \delta}q(x)\vert 0\rangle \Bigg\{\Big(\Gamma_{J}\Big)_{\rho \tau}
\nonumber\\
&\times &\text{Tr} \Big(  \tilde{\Delta}^{\lambda \delta}_{Q}(x) \gamma_\mu S^{(\text{pert.})}_{Q^{\prime}}(x) \gamma_\nu  \Big)
+
\Big( \Gamma_J \gamma_\nu \tilde{\Delta}^{\lambda \delta}_{Q}(x) \gamma_\mu S^{(\text{pert.})}_{Q^{\prime}}(x) \Big)_{\rho \tau}  \nonumber\\
&+&
\Big( \Gamma_J  \gamma_\nu   \tilde{S}^{(\text{pert.})}_{Q^{\prime}}(x) \gamma_\mu \Delta^{\lambda \delta}_{Q}(x) \Big)_{\rho \tau} 
+
\Big(  S^{(\text{pert.})}_{Q^{\prime}}(x)  \gamma_\nu \tilde{\Delta}^{\lambda \delta}_{Q}(x)  \gamma_\mu \Gamma_J  \Big)_{\rho \tau} 
\nonumber\\
&+&
\text{Tr}\Big(  \gamma_\nu \tilde{\Delta}^{\lambda \delta}_{Q}(x)  \gamma_\mu  \Gamma_J  \Big) \Big(S^{(\text{pert.})}_{Q^{\prime}}(x)\Big)_{\rho \tau}
-
\Big(S^{(\text{pert.})}_{Q^{\prime}}(x) \gamma_\nu \tilde{\Gamma}_J \gamma_\mu \Delta^{\lambda \delta}_{Q}(x)  \Big)_{\rho \tau}
\nonumber\\
&-&
\Big(  \Delta^{\lambda \delta}_{Q}(x)  \gamma_\nu   \tilde{S}^{(\text{pert.})}_{Q^{\prime}}(x) \gamma_\mu \Gamma_J  \Big)_{\rho \tau}
-
\Big(\Delta^{\lambda \delta}_{Q}(x)   \gamma_\nu \tilde{\Gamma}_J \gamma_\mu S^{(\text{pert.})}_{Q^{\prime}}(x)  \Big)_{\rho \tau}\nonumber
\\
&+&
\text{Tr} \Big(\Gamma_J  \gamma_\nu  \tilde{S}^{(\text{pert.})}_{Q^{\prime}}(x)  \gamma_\mu \Big) \Big(\Delta^{\lambda \delta}_{Q}(x)\Big)_{\rho \tau} \Bigg\},\nonumber
\\
\end{eqnarray}
where the superscript $(1)$ indicates one gluon exchange contribution. The contribution of exchanging gluon between the heavy quark $Q'$ and light vector meson can be found by simply exchanging $Q$ and $Q'$ in the above equation.

In calculation of the theoretical sides \ref{CFpert} and \ref{CFgluon}, one may run into some kinds of configurations which have the general forms of
\begin{eqnarray}\label{STR1}
T_{[~~,\alpha,\alpha\beta, \alpha\beta\gamma,...]}(p,q)&=& i \int d^4 x \int_{0}^{1} dv  \int {\cal D}\alpha e^{ip.x} \big(x^2 \big)^n  [e^{i (\alpha_{\bar q} + v \alpha _g) q.x} \mathcal{G}(\alpha_{i}) , e^{iq.x} f(u)] \nonumber\\
&\times& [1 , x_{\alpha} , x_{\alpha}x_{\beta},x_{\alpha}x_{\beta}x_{\gamma},...]  K_{n_1}(m_1\sqrt{-x^2})  K_{n_2}(m_2\sqrt{-x^2}),
\end{eqnarray} 
where the expressions in the brackets indicate different configurations resulting from calculations.
In the first bracket, the first and second expressions come from the two and three particle DAs respectively. The blank subscript on the LHS indicates no $x_{\alpha}$ in the corresponding configuration and the measure ${\cal D}\alpha$ is defined as 
\begin{equation*}
\int \mathcal{D}\alpha =\int_{0}^{1}d\alpha _{q}\int_{0}^{1}d\alpha _{\bar{q}%
}\int_{0}^{1}d\alpha _{g}\delta (1-\alpha _{q}-\alpha _{\bar{q}}-\alpha
_{g}).
\end{equation*}
Here we use the cosine representation of the Bessel function as follows:
\begin{equation}\label{CosineRep}
K_n(m_Q\sqrt{-x^2})=\frac{\Gamma(n+ 1/2)~2^n}{\sqrt{\pi}m_Q^n}\int_0^\infty dt~\cos(m_Qt)\frac{(\sqrt{-x^2})^n}{(t^2-x^2)^{n+1/2}}.
\end{equation}
There are several representations of the Bessel functions and it is shown in \cite{Azizi:2018duk} that the cosine representation makes the calculations straightforward and more suitable for the Borel transformation.
In this step, we perform the $x-$integration and  use 
\begin{eqnarray}\label{trick1}
(x^2)^n &=& (-1)^n \frac{d^n}{d \beta^n}\big(e^{- \beta x^2}\big)\arrowvert_{\beta = 0}, \nnb \\
x_{\alpha} e^{i P.x} &=& (-i) \frac{d}{d P^{\alpha}} e^{i P.x}.
\end{eqnarray}
After this stage, one needs to perform the double Borel transformation with respect to $(p+q)^2$ and  $p^2$ via
\begin{equation} \label{Borel1}
{\cal B}_{(p+q)^2}(M_{1}^{2}){\cal B}_{p^2}(M_{2}^{2})e^{b (p + u q)^2}=M^2 \delta(b+\frac{1}{M^2})\delta(u_0 - u) e^{\frac{-q^2}{M_{1}^{2}+M_{2}^{2}}},
\end{equation}
where  $u_0 = M_{1}^{2}/(M_{1}^{2}+M_{2}^{2})$ and $ \frac{1}{M^2}= \frac{1}{M_1^2}+\frac{1}{M_2^2}$. The details of the calculation can be found in \cite{Azizi:2018duk}.
For example, for the configuration 
\begin{eqnarray}\label{Z1}
{\cal Z}_{\alpha\beta}(p,q) &=& i \int d^4 x \int_{0}^{1} dv  \int {\cal D}\alpha e^{i[p+ (\alpha_{\bar q} + v \alpha _g)q].x} \mathcal{G}(\alpha_{i}) \big(x^2 \big)^n  \nonumber\\
&\times& x_\alpha x_\beta  K_{\mu}(m_Q\sqrt{-x^2})  K_{\nu}(m_Q\sqrt{-x^2}),
\end{eqnarray}
we get
 \begin{eqnarray}\label{STR4}
{\cal Z}_{\alpha\beta}(M^2) &=& \frac{i  \pi^2 2^{4-\mu-\nu} e^{\frac{-q^2}{M_1^2+M_2^2}}}{M^2 m_{Q_1}^{2\mu} m_{Q_2}^{2\nu}}\int  \mathcal{D}\alpha  \int_{0}^{1} dv \int_{0}^{1} dz \frac{\partial^n }{\partial \beta^n} e^{-\frac{m_1^2 \bar{z} + m_2^2 z}{z \bar{z}(M^2 - 4\beta)}} z^{\mu-1}\bar{z}^{\nu-1} (M^2 - 4\beta)^{\mu+\nu-1} \nonumber\\
&\times & \delta[u_0 - (\alpha_{q} + v \alpha_{g})]  \Big[ p_\alpha p_\beta + (v \alpha_{g} +\alpha_{q})(p_\alpha q_\beta +q_\alpha p_\beta ) + (v \alpha_{g} +\alpha_{q})^2 q_\alpha q_\beta \nonumber \\ 
&&+ \frac{M^2}{2}g_{\alpha\beta} \Big].
\end{eqnarray}

After completing the calculations in the hadronic and QCD sides in Borel scheme, we match the coefficients of the selected structures to obtain the desired sum rules as
\begin{eqnarray}
\label{eBQsBQV11}
g_1+g_2{m_2  \over m_1+m_2} \es {1\over 2 \lambda_{B_{Q_1}^*} \lambda_{B_{Q_2}^*}} 
e^{ {m_1^2\over M_1^2} + {m_2^2\over M_2^2} + {m_V^2\over M_1^2+M_2^2} }
\Pi_A~, \nnb \\
g_2 \es - {m_1+m_2 \over 2 \lambda_{B_{Q_1}^*} \lambda_{B_{Q_2}^*}}    
e^{ {m_1^2\over M_1^2} + {m_2^2\over M_2^2} + {m_V^2\over M_1^2+M_2^2} }
\Pi_B~, \nnb \\
g_3 \es - {(m_1+m_2)^2 \over \lambda_{B_{Q_1}^*} \lambda_{B_{Q_2}^*}}  
e^{ {m_1^2\over M_1^2} + {m_2^2\over M_2^2} + {m_V^2\over M_1^2+M_2^2} }
\Pi_C~, \nnb \\
g_4 \es - {(m_1+m_2)^3 \over 2 \lambda_{B_{Q_1}^*} \lambda_{B_{Q_2}^*}}                     
e^{ {m_1^2\over M_1^2} + {m_2^2\over M_2^2} + {m_V^2\over M_1^2+M_2^2} }
\Pi_D~,
\end{eqnarray}
where the explicit forms of  $\Pi_A$, $\Pi_B$, $\Pi_C$ and $\Pi_D$  functions   are presented in the Appendix \ref{APA}. As the mass of the initial and final baryons are the same/close, we set $M_1^2 = M_2^2 = 2 M^2$ and $ u_0=1/2 $. To further suppress the contributions of the higher states and continuum  we perform the subtraction procedure according to the standard prescriptions of the method. In this case,  it is enough to set
\begin{eqnarray}
  e^{-\frac{m_1^2 \bar{z} + m_2^2 z}{M^2 z \bar{z}}} = e^{-s_0/M^2},
\end{eqnarray}
where $s_0$ is the continuum threshold. This leads to
\begin{eqnarray}\label{zsubtraction}
\int_{0}^{1}dz \rightarrow \int_{z_{\text{min}}}^{z_{\text{max}}}dz,
\end{eqnarray}
where $z_{\text{max}}$ and $z_{\text{min}}$ are  given as
\begin{eqnarray}\label{zlimits}
z_{\text{max}(\text{min})}=\frac{1}{2s_0}\Big[(s_0+m_1^2-m_2^2)+(-)\sqrt{(s_0+m_1^2-m_2^2)^2-4m_1^2s_0}\Big].
\end{eqnarray}

\section{Numerical analyses}\label{NA}
In this section, we will go over  the numerical analyses of the sum rules for the strong coupling constants of the spin-$3/2$ doubly heavy baryons  with light  vector mesons. To this end, two sets of input parameters are needed. The first one corresponds to nonperturbative parameters coming from distribution amplitudes (DAs), associated with different twists normalized to the mass
scale $ \mu=1$~GeV as well as the mass and decay constants of the light vector mesons.
These parameters are gathered in tables \ref{tabmass} and \ref{tabf}.
The  second set corresponds to the masses and residues of the doubly heavy hadrons with spin-$3/2$. 
These parameters are taken from  \cite{Aliev:2012iv}  and are listed in table \ref{tabBaryon}.

\begin{table}[t]
	\renewcommand{\arraystretch}{1.3}
	\addtolength{\arraycolsep}{1pt}
	$$
	\rowcolors{2}{purple!10}{white}
	\begin{tabular}{|c|c|c|c|}
	\rowcolor{purple!30}
	\hline \hline
	\mbox{Parameters}         &  \mbox{Values  }  
	\\
	\hline\hline
		$ m_s $    & $  95^{+9}_{-3}~\mbox{MeV} $      \\
	$ m_{c} $    &  $ 1.275^{+0.025}_{-0.035}~\mbox{GeV} $      \\
	$ m_b $    & $  4.18^{+0.04}_{-0.03}~\mbox{GeV} $      \\
	$ m_{\rho } $    &   $775.26\pm0.25 ~\mbox{MeV}$  \\
	$ m_{\omega} $    &   $782.65\pm0.12~\mbox{MeV} $  \\
		$ m_{\phi} $    &   $1019.461\pm0.016 ~\mbox{MeV}$  \\
	$ m_{K^{*0}}    $  &   $895.55\pm0.020~\mbox{MeV}$    \\
		$ m_{\bar{K}^{*0}}   $  &   $895.55\pm0.020~\mbox{MeV}$    \\
	$ m_{K^{*\pm}}  $  &   $891.76 \pm0.025~\mbox{MeV}$    \\
	\hline \hline
	\end{tabular}
	$$
	\caption{The masses of quarks and vector  mesons \cite{Tanabashi:2018oca}.}  \label{tabmass} 
	\renewcommand{\arraystretch}{1}
	\addtolength{\arraycolsep}{-1.0pt}
\end{table}

\begin{table}[t]
	\renewcommand{\arraystretch}{1.3}
	\addtolength{\arraycolsep}{1pt}
	$$
	\rowcolors{2}{purple!10}{white}
	\begin{tabular}{|c|c|c|c|}
	\rowcolor{purple!30}
	\hline \hline
	\mbox{Parameters}         &  \mbox{Values  $[\text{MeV}]$ }  
	\\
	\hline\hline
	$  f_\rho^\parallel $    & $  216\pm3$     \\
	$  f_\omega^\parallel $    & $  187\pm5  $     \\
	$  f_{K^*}^\parallel $    & $  220\pm5  $     \\
	$  f_\phi^\parallel$    & $  215\pm5  $     \\
		$  f_\rho^\perp $    & $  165\pm9  $     \\
	$  f_\omega^\perp $    & $  151\pm9  $     \\
	$  f_{K^*}^\perp $    & $  185\pm10  $     \\
	$  f_\phi^\perp$    & $  186\pm9  $     \\
	\hline \hline
	\end{tabular}
	$$
	\caption{The longitudinal ($ f^\parallel  $) and transverse ($ f^\perp $) decay constants  for the vector mesons $\rho $,  $\omega $, and $\phi $  \cite{Ball:2007rt} }.  \label{tabf} 
	\renewcommand{\arraystretch}{1}
	\addtolength{\arraycolsep}{-1.0pt}
\end{table}

\begin{table}[t]
	\renewcommand{\arraystretch}{1.3}
	\addtolength{\arraycolsep}{1pt}
	$$
	\rowcolors{2}{purple!10}{white}
	\begin{tabular}{|c|c|c|c|}
	\rowcolor{purple!30}
	\hline \hline
	\mbox{Baryon}         &  \mbox{Mass} $[\text{GeV}]$  & \mbox{Residue $[\text{GeV}^3]$}
	\\
	\hline\hline
	
	$ \Xi^*_{cc}  $    &  $ 369\pm0.16 $  & $0.12\pm0.01$ \\
	$ \Xi^*_{bc}  $    &  $ 7.25\pm0.20 $ & $0.15\pm0.01$  \\
	$ \Xi^*_{bb} $    &  $ 10.14\pm1.0 $  & $0.22\pm0.03$  \\
	$ \Omega^*_{cc}  $    &  $ 3.78\pm0.16 $  & $0.14\pm0.02$ \\
	$ \Omega^*_{bc}  $    &  $ 7.3\pm0.2 $  & $0.18\pm0.02$ \\
	$ \Omega^*_{bb} $    &  $ 10.5\pm0.2  $  & $0.25\pm0.03$ \\
	
	\hline \hline
\end{tabular}
$$
\caption{The  masses and residues of spin-$ 3/2 $ doubly heavy baryons  \cite{Aliev:2012iv}.}  \label{tabBaryon} 
\renewcommand{\arraystretch}{1}
\addtolength{\arraycolsep}{-1.0pt}
\end{table} 
 
The continuum threshold $ s_0 $ and the Borel parameter $ M^2 $ are the auxiliary parameters of
the sum rules for the strong coupling constants that are needed to be fixed.  Consequently, we need to locate the working region of these parameters, where the variations of the results concerning the changes in these parameters should be minimal.

The continuum threshold $s_0$  depends on the energy of the first excited state at each channel. There is no experimental information on the first excited states in the  doubly heavy baryon channels, hence, we pick up the region $(m_{B^*}+0.3)^2\leq s_0\leq(m_{B^*}+0.7)^2~ \text{GeV}^2$, where the variation of the results with respect to the threshold parameter are mild. Figs.~\ref{fig:s0},  \ref{fig:s0bc} and \ref{fig:s0bb}, as examples, display  the dependence of the strong coupling constants $ g_1$, $ g_2$, $ g_3 $ and
$ g_4 $ on the parameter $ s_0 $ for the vertices $\Xi^*_{cc} \Xi^*_{cc} \rho^\pm(\omega)  $, $\Xi^*_{bc} \Xi^*_{bc} \rho^\pm(\omega)  $  and $\Xi^*_{bb} \Xi^*_{bb} \rho^\pm(\omega)$ respectively.
These figures  reveal that the values of the strong coupling constants  remain roughly unchanged at the presented region of $ s_0 $.

In this step, we would like to fix the working region of  the Borel parameter $ M^2 $.
The lower limit of $  M^2$ is determined by the requirement of the OPE convergence.
The upper limit of this parameter is obtained  demanding the pole dominance, i.e., the pole contribution should be more than  $ 50\% $ of the total contribution for each of the selected structures. As a result of these requirements the Borel windows are determined to be  $14$~GeV$ ^2  $ $\leq M^2$ $\leq 18$~GeV $^2 $, $8$~GeV$ ^2  $ $\leq M^2$ $\leq 11$~GeV $^2 $ and $4$~GeV$ ^2  $ $\leq M^2$ $\leq 6$~GeV $^2 $ for $ bb $, $ bc $ and $ cc $ channels, respectively.
Figs.~\ref{fig:msq}, \ref{fig:msqbc} and \ref{fig:msqbb}, as examples, illustrate the dependence of the strong coupling constants  $ g_1$, $ g_2$, $ g_3 $ and$ g_4 $  on the Borel mass $ M^2 $ for the vertices $\Xi^*_{cc} \Xi^*_{cc} \rho^\pm (\omega ) $,  $\Xi^*_{bc} \Xi^*_{bc} \rho^\pm(\omega)  $  and $\Xi^*_{bb} \Xi^*_{bb} \rho^\pm(\omega)$, respectively at average values of the continuum threshold.  Our analyses show that  the variations of the strong coupling constants with respect to  $ M^2 $  are relatively small in the selected Borel window. 

Using all the input values as well as the working intervals for the auxiliary parameters we find the numerical values for the strong coupling constants under study from the obtained sum rules for the strong coupling form factors at $q^2 = m_V^2$. The predicted results  for $ g_1$, $ g_2$, $ g_3 $ and
$ g_4 $  for all  the considered vertices in $ bb $, $ bc $ and $ cc $ channels are  demonstrated in Table~\ref{tabmeson}. The errors in the values can be attributed to the uncertainties in the calculations of   the working regions for  $ M^2 $ and $ s_0 $, the errors in the masses and
residues of the doubly heavy baryons and the uncertainties originating from the DAs parameters.
Our results may be checked via different nonperturbative approaches.

\begin{figure}[h!]
\includegraphics[width=1.0\textwidth]{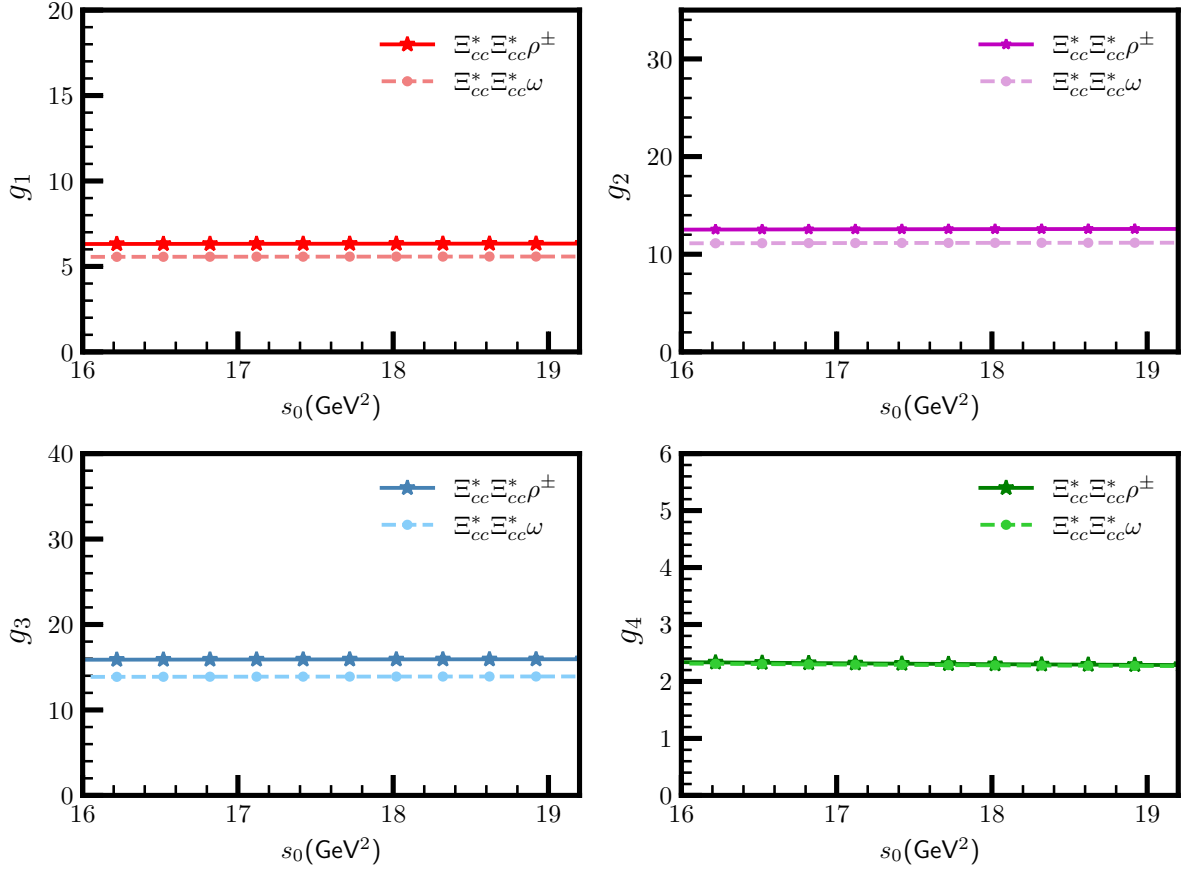}
\caption{The strong couplings $g_i$ 
		as functions of continuum threshold $s_0$ for the vertices $\Xi^*_{cc} \Xi^*_{cc} \rho^\pm  $  and $\Xi^*_{cc} \Xi^*_{cc} \omega  $ at $ M^2=5 $~GeV$ ^2 $.}
\label{fig:s0}
\end{figure} 

\begin{figure}[h!]
	\includegraphics[width=1.0\textwidth]{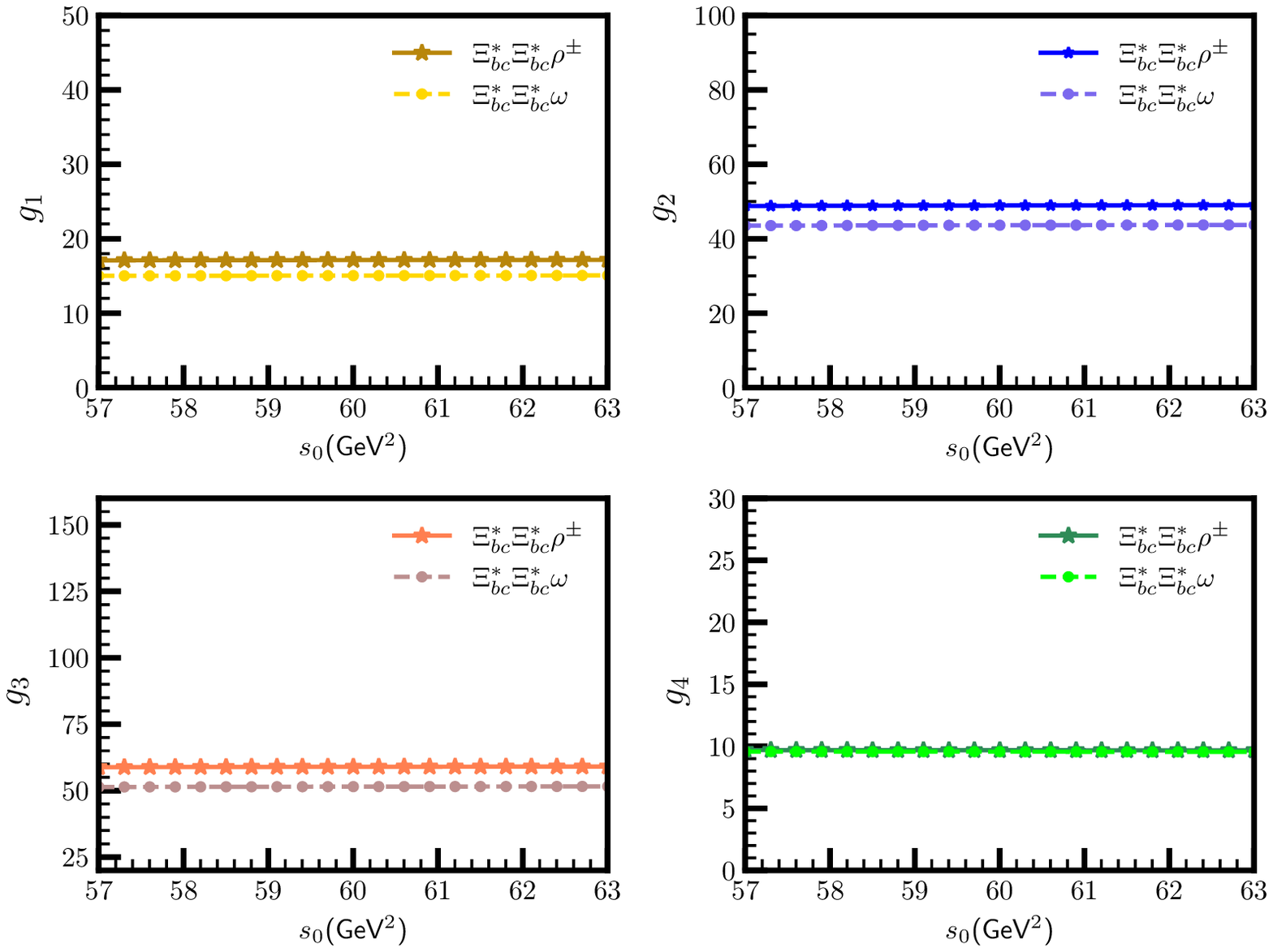}
	\caption{The strong couplings $g_i$ 
		as functions of continuum threshold $s_0$ for the vertices $\Xi^*_{bc} \Xi^*_{bc} \rho^\pm  $  and $\Xi^*_{bc} \Xi^*_{bc} \omega  $ at $ M^2=9.5 $~GeV$ ^2 $.}
	\label{fig:s0bc}
\end{figure} 

\begin{figure}[h!]
	\includegraphics[width=1.0\textwidth]{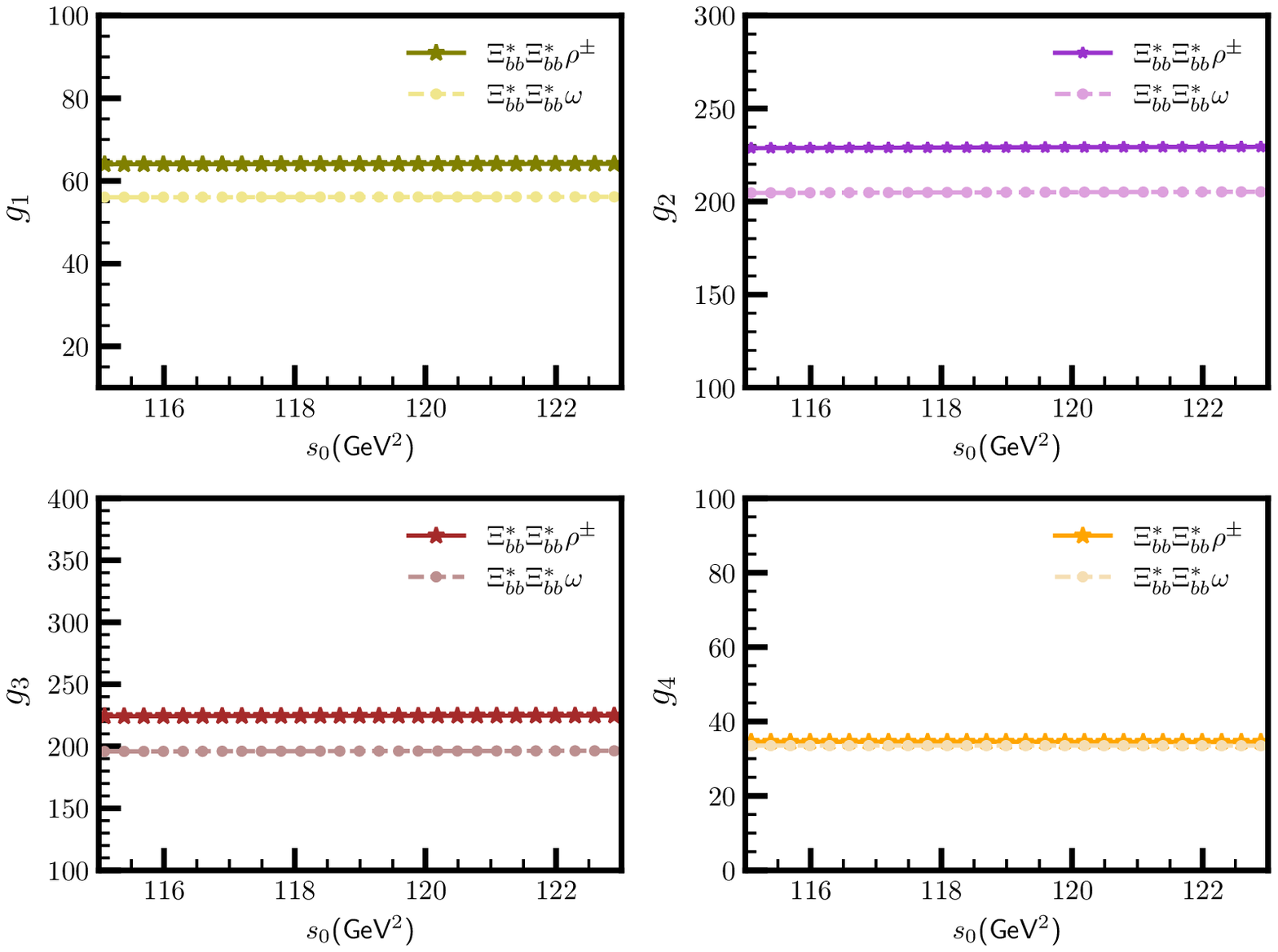}
	\caption{The strong couplings $g_i$ 
		as functions of continuum threshold $s_0$ for the vertices $\Xi^*_{bb} \Xi^*_{bb} \rho^\pm  $  and $\Xi^*_{bb} \Xi^*_{bb} \omega  $ at $ M^2=16 $~GeV$ ^2 $.}
	\label{fig:s0bb}
\end{figure} 

\begin{figure}[h!]
	\includegraphics[width=1.0\textwidth]{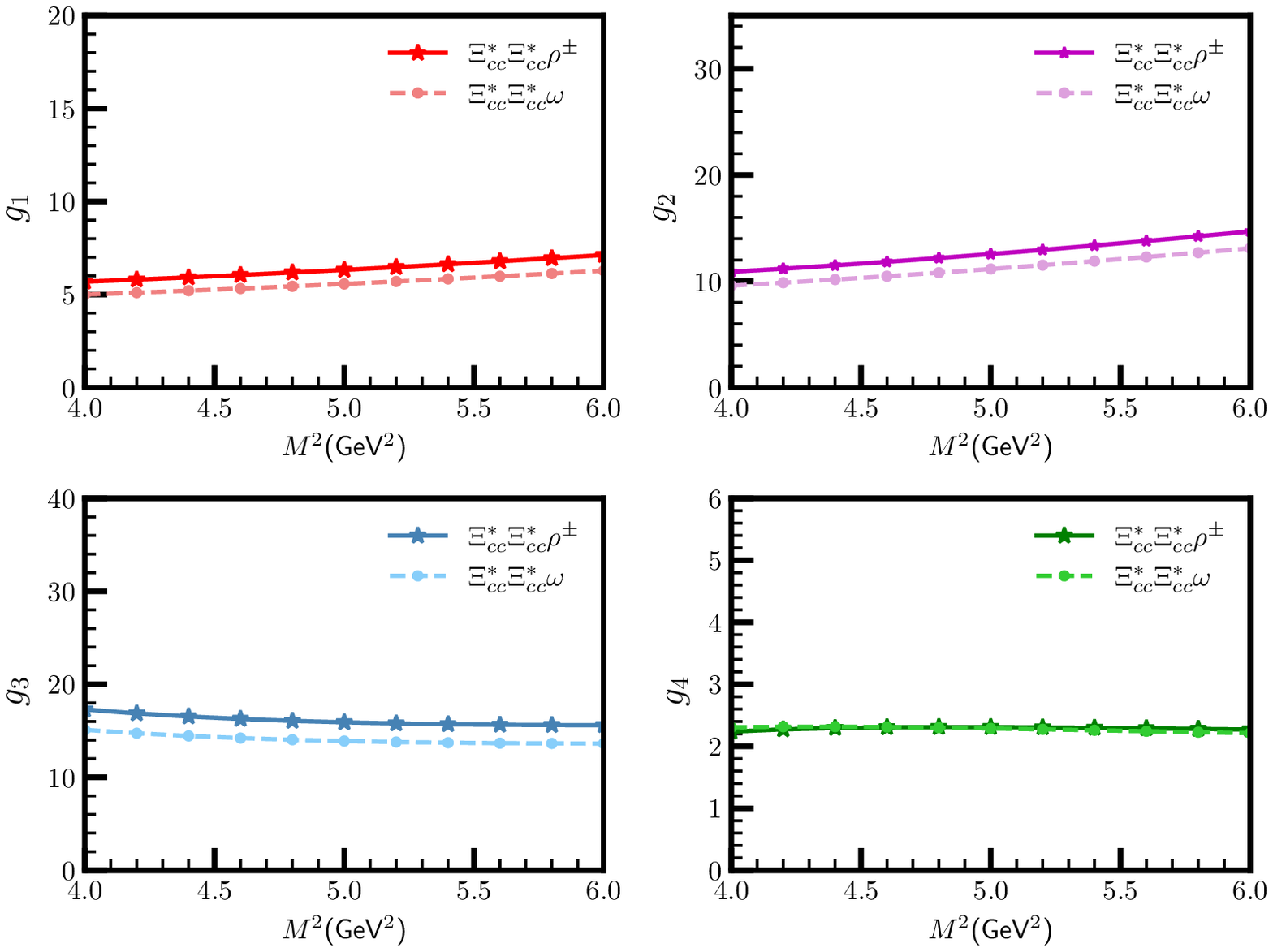}
	\caption{The strong couplings $g_i$ 
		as functions of the Borel
		parameter $ M^2 $ for the vertices $\Xi^*_{cc} \Xi^*_{cc} \rho^\pm  $  and $\Xi^*_{cc} \Xi^*_{cc} \omega  $ at  $ s_0=17.6 $~GeV$ ^2 $.}
	\label{fig:msq}
\end{figure}

\begin{figure}[h!]
	\includegraphics[width=1.0\textwidth]{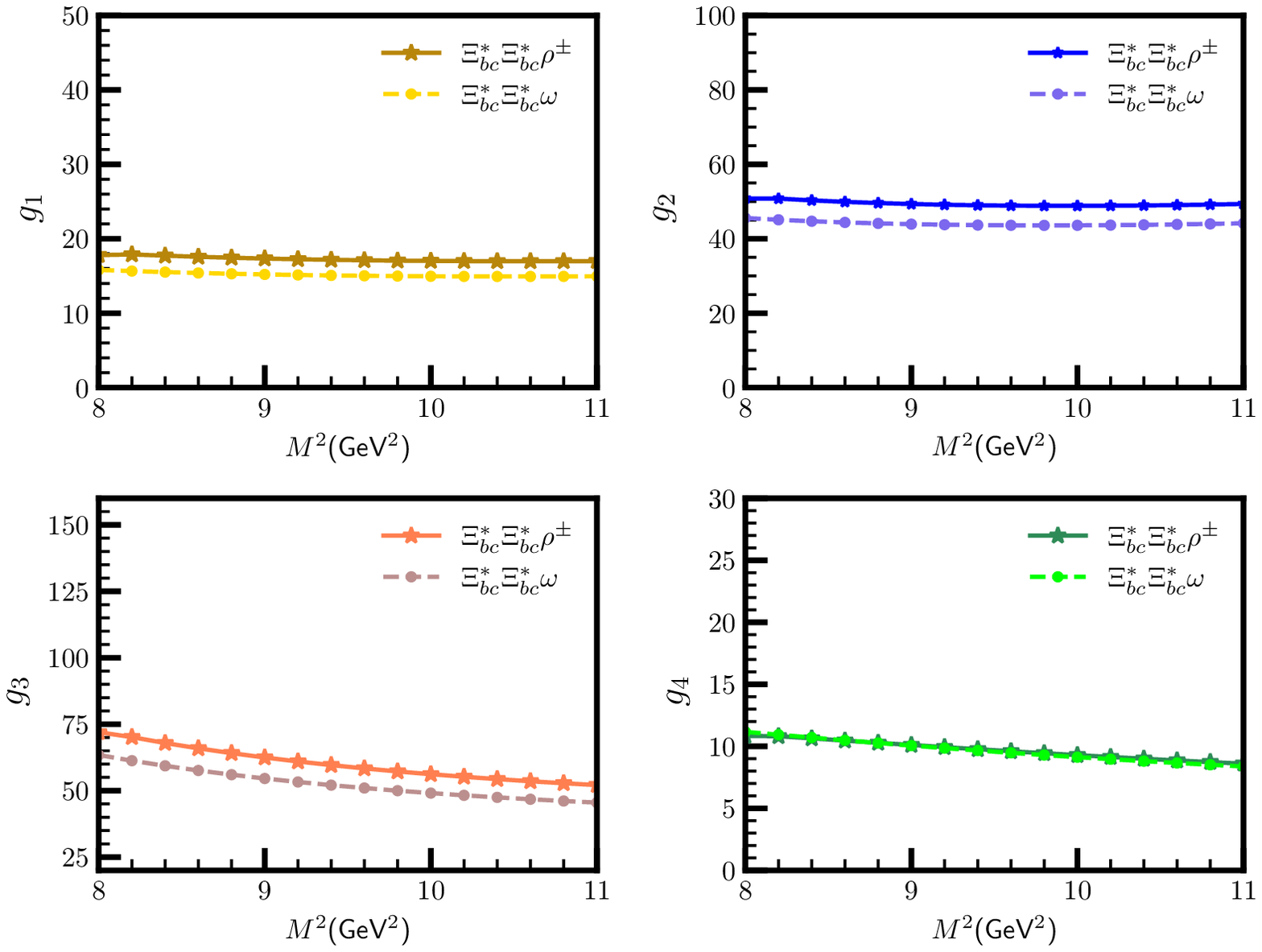}
	\caption{The strong couplings $g_i$ 
		as functions of the Borel
		parameter $ M^2 $ for the vertices $\Xi^*_{bc} \Xi^*_{bc} \rho^\pm  $  and $\Xi^*_{bc} \Xi^*_{bc} \omega  $ at  $ s_0=60.1 $~GeV$ ^2 $.}
	\label{fig:msqbc}
\end{figure}

\begin{figure}[h!]
	\includegraphics[width=1.0\textwidth]{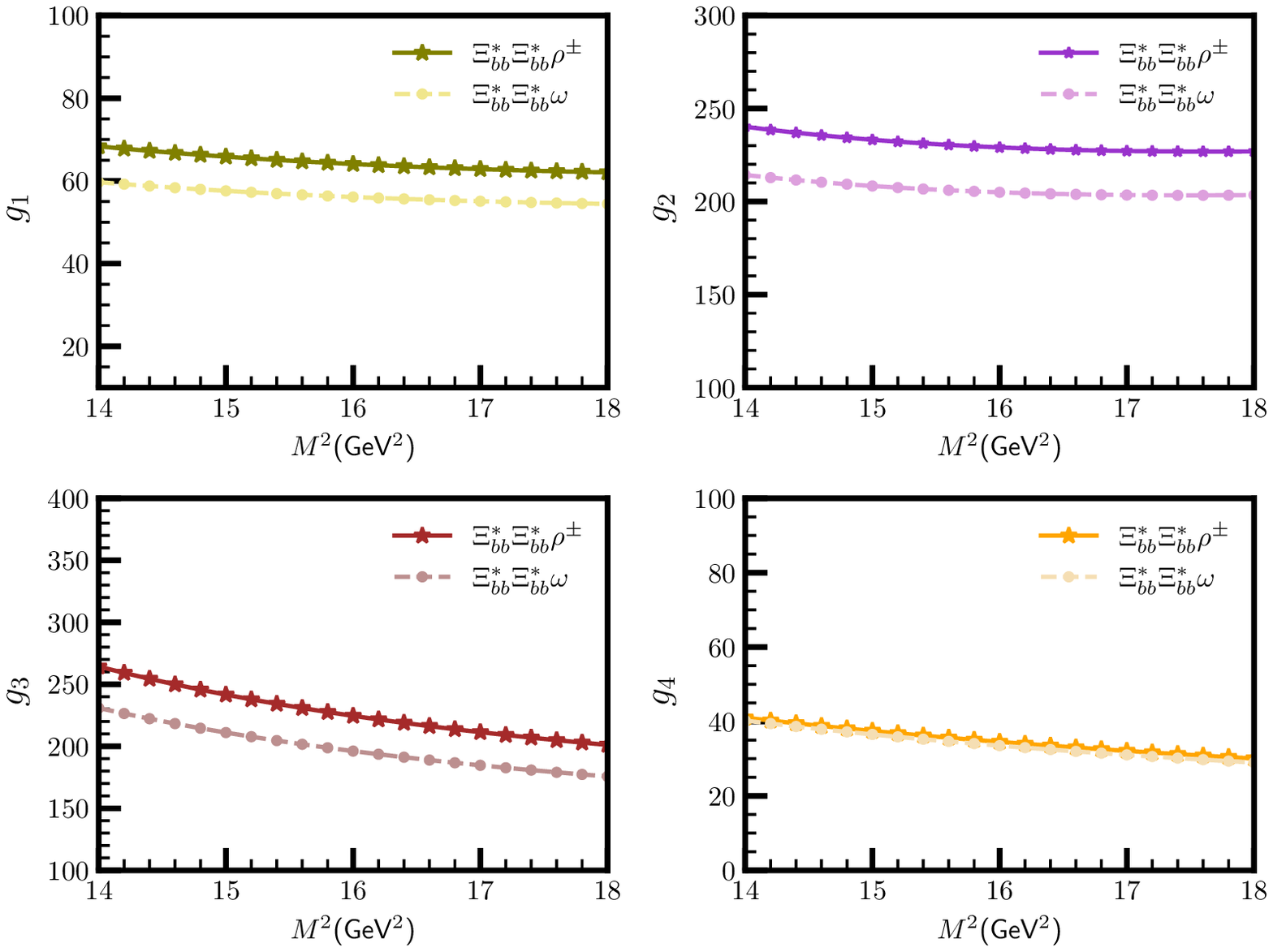}
	\caption{The strong couplings $g_i$ 
		as functions of the Borel
		parameter $ M^2 $ for the vertices $\Xi^*_{bb} \Xi^*_{bb} \rho^\pm  $  and $\Xi^*_{bb} \Xi^*_{bb} \omega  $ at  $ s_0=118.9 $~GeV$ ^2 $.}
	\label{fig:msqbb}
\end{figure}

\begin{table}[t]
	\renewcommand{\arraystretch}{1.3}
	\addtolength{\arraycolsep}{1pt}
	$$
	\begin{tabular}{|c|c|c|c|c|}
	\rowcolor{purple!30}
	\hline \hline
	\mbox{Vertex} & $ g_1 $   &$ g_2 $  &$ g_3 $  &  $ g_4  $
	\\
		\hline \hline
			\rowcolor{purple!15}
	$ \Xi^*_{bb}\Xi^*_{bb} \rho^0 $   &  $  45.8^{\:2.5}_{\:2.0}$ & $  164.1^{\:5.9}_{\:2.0}$  &  $  162.6^{\:24}_{\:20}$  &   $ 24.8^{\:4.1}_{\:3.6}$   \\
		\rowcolor{purple!5}
	$ \Xi^*_{bc}\Xi^*_{bc} \rho^0 $   & $  12.3^{\:0.4}_{\:0.3}$  &  $  35.3^{\:1.0}_{\:0.7}$ & $  43.3^{\:8.1}_{\:6.5}$  &  $ 6.9^{\:0.8}_{\:0.8}$    \\
	\rowcolor{purple!1}
    $ \Xi^*_{cc}\Xi^*_{cc} \rho^0 $   &  $  4.5^{\:0.5}_{\:0.4}$ & $  8.98^{\:1.4}_{\:1.3}$  &  $  11.5^{\:0.7}_{\:0.5}$  &$ 1.6^{\:0.04}_{\:0.04}$    \\
	\hline \hline
				\rowcolor{purple!15}
	$ \Xi^*_{bb}\Xi^*_{bb} \rho^\pm $   &  $  64.8^{\:3.5}_{\:2.8}$ & $  232.04^{\:8.4}_{\:5.6}$  &  $  229.9^{\:34.5}_{\:29.2}$  &   $  35.2^{\:5.8}_{\:5.2}$   \\
		\rowcolor{purple!5}
	$ \Xi^*_{bc}\Xi^*_{bc} \rho^\pm $   & $  17.4^{\:0.6}_{\:0.4}$  &  $  49.8^{\:1.5}_{\:1.1}$ & $  61.2^{\:11.4}_{\:9.3}$  &  $  9.7^{\:1.2}_{\:1.1}$    \\
	\rowcolor{purple!1}
    $ \Xi^*_{cc}\Xi^*_{cc} \rho^\pm $   &  $  6.3^{\:0.7}_{\:0.6}$ & $  12.7^{\:2}_{\:1.8}$  &  $  16.3^{\:1.0}_{\:0.7}$  &$  2.3^{\:0.06}_{\:0.06}$    \\
	\hline \hline
		\rowcolor{purple!15}
		$ \Xi^*_{bb}\Xi^*_{bb} \omega $   &  $ 56.7 ^{\:2.9}_{\:2.3}$ & $  207.5^{\:6.9}_{\:4.5}$  &  $  200.9^{\:30.1}_{\:25.5}$  &   $ 34.2 ^{\:6.0}_{\:5.2}$  \\
		\rowcolor{purple!5}
	$ \Xi^*_{bc}\Xi^*_{bc} \omega $  &  $  15.27^{\:2.9}_{\:2.3}$ & $  44.44^{\:1.1}_{\:0.9}$  &  $  53.46^{\:10}_{\:8}$  &  $  9.70^{\:1.4}_{\:1.3}$   \\
	\rowcolor{purple!1}
    $ \Xi^*_{cc}\Xi^*_{cc} \omega $   &  $  5.6^{\:0.7}_{\:0.6}$ & $  11.3^{\:1.8}_{\:1.7}$  &  $  14.2^{\:0.9}_{\:0.6}$  &   $  2.2^{\:0.06}_{\:0.07}$   \\
	\hline \hline
	\rowcolor{purple!15}
			$ \Omega^*_{bb}\Omega^*_{bb} \phi $   &  $  62.7^{\:1.6}_{\:1.2}$ & $  215.5^{\:7.2}_{\:4.8}$  &  $  264.9^{\:44}_{\:37}$  &   $  33.1^{\:3.3}_{\:3.2}$  \\
			\rowcolor{purple!5}
	$\Omega^*_{bc} \Omega^*_{bc} \phi$   &  $  14.2^{\:0.2}_{\:0.1}$ & $  37.3^{\:0.5}_{\:0.6}$  &  $  60.2^{\:12.1}_{\:9.8}$  &  $  5.2^{\:0.7}_{\:1.2}$  \\
	\rowcolor{purple!1}
    $ \Omega^*_{cc} \Omega^*_{cc} \phi $   &  $  6.5^{\:0.8}_{\:0.8}$ & $  10.1^{\:1.9}_{\:1.8}$  &  $  18.5^{\:1.7}_{\:1.2}$  & $ 1.5^{\:0.4}_{\:0.6}$  \\
	\hline \hline
	\rowcolor{purple!15}
				$ \Omega^*_{bb} \Xi^*_{bb} K^{*0} $   &  $ 71.5^{\:3.3}_{\:2.6}$ & $  251.8^{\:9}_{\:6}$  &  $  262.8^{\:4}_{\:3}$  &   $  25.3^{\:1.6}_{\:1.7}$ \\
				\rowcolor{purple!5}
	$ \Omega^*_{bc} \Xi^*_{bc}  K^{*0}  $   &  $  18.1^{\:0.4}_{\:0.2}$ & $  49.0^{\:1.0}_{\:0.9}$  &  $  64.7^{\:12}_{\:10}$  &    $  3.9^{\:1.0}_{\:1.6}$\\
	\rowcolor{purple!1}
    $\Omega^*_{cc} \Xi^*_{cc}  K^{*0}   $   &  $  7.4^{\:0.8}_{\:0.7}$ & $  13.1^{\:2.0}_{\:2.0}$  &  $  18.5^{\:1.4}_{\:1.0}$  &   $  1.0^{\:0.5}_{\:0.6}$ \\
	\hline \hline
		\rowcolor{purple!15}
				$ \Xi^*_{bb} \Omega^*_{bb} \bar{K}^{*0} $   &  $ 72.7^{\:3.4}_{\:2.7}$ & $  251.6^{\:9.1}_{\:6.2}$  &  $  262.6^{\:41}_{\:35}$  &   $  25.3^{\:1.6}_{\:1.7}$ \\
				\rowcolor{purple!5}
	$ \Xi^*_{bc} \Omega^*_{bc}  \bar{K}^{*0} $   &    $ 18.2^{\:0.4}_{\:0.2}$ & $  49.0^{\:1.1}_{\:0.9}$  &  $  64.7^{\:12.5}_{\:10.2}$  &   $  4.0^{\:1.0}_{\:1.6}$ \\
	\rowcolor{purple!1}
    $\Xi^*_{cc} \Omega^*_{cc} \bar{K}^{*0}$   &    $ 7.6^{\:0.8}_{\:0.8}$ & $  13.1^{\:2.1}_{\:2.0}$  &  $  18.4^{\:1.4}_{\:1.0}$  &   $  1.1^{\:0.5}_{\:0.6}$ \\
	\hline \hline
	\end{tabular}
	$$
	\caption{Strong coupling costants of the doubly heavy  spin-$ 3/2 $ baryons with  light vector mesons obtained from the sum rules at $q^2 = m_V^2$.}  \label{tabmeson} 
	\renewcommand{\arraystretch}{1}
	\addtolength{\arraycolsep}{-1.0pt}
\end{table}

\newpage
\section{Summary and conclusions}\label{SC}

After the observation of $\Xi_{cc}^{++}$ by the  LHCb collaboration, the possibility for identification of other doubly heavy baryons predicted by the quark model has  increased. In parallel, many theoretical studies on different properties of doubly heavy baryons are performed. One of the important issues in hadron physics is to determine the strong coupling constants among different hadronic multiplets. In this connection, we studied the strong vertices of the spin-3/2 doubly heavy baryons with light vector $\rho$, $K^*$, $\omega$ and $\phi$ mesons. In the calculations that were done via LCSR, we used the interpolating currents of the baryons and DAs of the vector mesons. The correlation function in spin-3/2 channel receives contributions also from the spin-1/2 doubly heavy baryons, that were removed by an appropriate ordering of Dirac matrices and selection of Lorentz structure free from the unwanted contributions.  Such vertices are parameterized in terms of four  strong coupling form factors. The values of these strong form factors  at $q^2 = m_V^2$ are the strong coupling form factors, which have been determined.  These strong coupling form factors are basic quantities that carry information not only on the internal structure of the participating particles but also on the nature of strong interactions among the doubly heavy baryons and vector mesons. The values of these coupling constants may be used in the determination of the strong potentials among the hadronic multiplets. We hope that we will have more experimental data on different physical quantities of the doubly heavy baryons. Our results may help experimental groups in analyses of the related data.

\section*{ACKNOWLEDGEMENTS}
K. Azizi and S. Rostami are thankful to Iran Science Elites Federation (Saramadan)
for the financial support provided under the grant number ISEF/M/99171.

\clearpage
\newpage
\section*{Appendix A:\\ Invariant  Functions $\Pi_A$, $\Pi_B$, $\Pi_C$ and $\Pi_D$. } \label{APA}
In this appendix, we  give  the explicit forms of the $\Pi_A$, $\Pi_B$, $\Pi_C$ and $\Pi_D$ functions. Each $\Pi_i$ function  can be written  as
\begin{eqnarray}
\Pi_i = \Pi_i^0 + \Pi_i^{GG},
\end{eqnarray}
where the components $ \Pi_{A,B,CD}^0 $   and  $\Pi_{A,B,CD}^{GG}  $ for the  $\Omega^*_{cc}\Xi^*_{cc}K^*$ vertex (as an example) are given as:
\begin{eqnarray}
\Pi_A^{0}&=&\frac{ e^{-\frac{q^2}{M_1^2+M_2^2}}}{24 M^2 \pi ^2} \Biggl\{m_c M^2 {\cal A}^\perp(u_0) f_\rho^\perp m_\rho^2
(-m_c^2 \zeta _{-1,0}(m_c)-m_c^2 \zeta _{0,-1}(m_c)+m_c^2 \zeta _{0,0}(m_c)\nonumber\\
&-&2 M^2 \zeta _{0,0}(m_c)+m_c^2
\zeta _{1,-1}(m_c)+2 M^2 \zeta _{1,0}(m_c))+2 m_c M^2 f_\rho^\perp \Bigg[2 M^4 \phi^\perp(u_0)
(\zeta _{0,0}(m_c)-\zeta _{1,0}(m_c))\nonumber\\
&+&m_\rho^2 \Bigg(-6 q^2 u_0 i_1({\cal T}_4(\alpha_i),1) \zeta
_{-1,0}(m_c)-M^2 i_2({\cal S}(\alpha_i),1) \zeta _{-1,0}(m_c)+3 M^2 i_2(\tilde{\cal S}(\alpha_i),1) \zeta _{-1,0}(m_c)\nonumber\\
&-&M^2
i_2({\cal T}_3(\alpha_i),1) \zeta _{-1,0}(m_c)-M^2 i_2({\cal T}_4(\alpha_i),1) \zeta _{-1,0}(m_c)\nonumber\\
&+&4 q^2 u_0
i_1({\cal T}(\alpha_i),1) (\zeta _{-1,0}(m_c)-\zeta _{0,0}(m_c))-16 M^2 u_0 \psi^\parallel(u_0) \zeta
_{0,0}(m_c)\nonumber\\
&+&4 q^2 u_0 i_1({\cal T}_4(\alpha_i),1) \zeta _{0,0}(m_c)-5 M^2 i_2({\cal S}(\alpha_i),1) \zeta
_{0,0}(m_c)\nonumber\\
&+&8 M^2 i_2({\cal S}(\alpha_i),v) \zeta _{0,0}(m_c)-3 M^2 i_2(\tilde{S}(\alpha_i),1) \zeta _{0,0}(m_c)+M^2
i_2({\cal T}_1(\alpha_i),1) \zeta _{0,0}(m_c)\nonumber\\
&-&2 M^2 i_2({\cal T}_1(\alpha_i),v) \zeta _{0,0}(m_c)-M^2 i_2({\cal T}_2(\alpha_i),1)
\zeta _{0,0}(m_c)+2 M^2 i_2({\cal T}_2(\alpha_i),v) \zeta _{0,0}(m_c)\nonumber\\
&+&2 M^2 i_2({\cal T}_3(\alpha_i),1) \zeta _{0,0}(m_c)-2
M^2 i_2({\cal T}_3(\alpha_i),v) \zeta _{0,0}(m_c)+M^2 i_2({\cal T}_4(\alpha_i),1) \zeta _{0,0}(m_c)\nonumber\\
&+&2 M^2
i_2({\cal T}_4(\alpha_i),v) \zeta _{0,0}(m_c)+6 q^2 u_0 i_1({\cal T}_3(\alpha_i),1) (-\zeta _{-1,0}(m_c)+\zeta
_{0,0}(m_c))\nonumber\\
&+&16 M^2 u_0 \psi^\parallel(u_0) \zeta _{1,0}(m_c)\Bigg)\Bigg]+2 f_\rho^\parallel m_\rho
\Bigg[m_\rho^2 \Bigg(12 m_c^2 M^2 i_1(\tilde{\phi}(\alpha_i),1) \zeta _{-1,0}(m_c)\nonumber\\
&-&4 m_c^2 M^2 i_1(\psi(\alpha_i),1)
\zeta _{-1,0}(m_c)+8 m_c^2 q^2 u_0^2 i_1(\psi(\alpha_i),1) \zeta _{-1,0}(m_c)+8 m_c^2 M^2
i_1(\psi(\alpha_i),v) \zeta _{-1,0}(m_c)\nonumber\\
&-&16 m_c^2 q^2 u_0^2 i_1(\psi(\alpha_i),v) \zeta _{-1,0}(m_c)-8
m_c^2 q^2 u_0 i_1(\alpha_g \psi(\alpha_i),v) \zeta _{-1,0}(m_c)\nonumber\\
&+&16 m_c^2 q^2 u_0 i_1(\alpha_g
\psi(\alpha_i),v^2) \zeta _{-1,0}(m_c)-8 m_c^2 q^2 u_0 i_1(\alpha_q \psi(\alpha_i),1) \zeta _{-1,0}(m_c)\nonumber\\
&+&16
m_c^2 q^2 u_0 i_1(\alpha_q \psi(\alpha_i),v) \zeta _{-1,0}(m_c)+12 m_c^2 M^2 i_1(\tilde{\psi}(\alpha_i),1)
\zeta _{-1,0}(m_c)\nonumber\\
&+&40 m_c^2 q^2 u_0^2 i_1(\tilde{\psi}(\alpha_i),1) \zeta _{-1,0}(m_c)-40 m_c^2 q^2
u_0 i_1(\alpha_g \tilde{\psi}(\alpha_i),v) \zeta _{-1,0}(m_c)\nonumber\\
&-&40 m_c^2 q^2 u_0 i_1(\alpha_q \tilde{\psi}(\alpha_i),1)
\zeta _{-1,0}(m_c)+3 m_c^4 j_1({\cal A}^\parallel (u)) \zeta _{-1,0}(m_c)+4 m_c^2 M^2 i_1(\tilde{\phi}(\alpha_i),1)
\zeta _{0,-1}(m_c)\nonumber\\
&-&4 m_c^2 M^2 i_1(\psi(\alpha_i),1) \zeta _{0,-1}(m_c)+8 m_c^2 q^2 u_0^2
i_1(\psi(\alpha_i),1) \zeta _{0,-1}(m_c)+8 m_c^2 M^2 i_1(\psi(\alpha_i),v) \zeta _{0,-1}(m_c)\nonumber\\
&-&16 m_c^2
q^2 u_0^2 i_1(\psi(\alpha_i),v) \zeta _{0,-1}(m_c)-8 m_c^2 q^2 u_0 i_1(\alpha_g \psi(\alpha_i),v)
\zeta _{0,-1}(m_c)\nonumber\\
&+&16 m_c^2 q^2 u_0 i_1(\alpha_g \psi(\alpha_i),v^2) \zeta _{0,-1}(m_c)-8
m_c^2 q^2 u_0 i_1(\alpha_q \psi(\alpha_i),1) \zeta _{0,-1}(m_c)\nonumber\\
&+&16 m_c^2 q^2 u_0 i_1(\alpha_q
\psi(\alpha_i),v) \zeta _{0,-1}(m_c)+4 m_c^2 M^2 i_1(\tilde{\psi}(\alpha_i),1) \zeta _{0,-1}(m_c)\nonumber\\
&+&24 m_c^2
q^2 u_0^2 i_1(\tilde{\psi}(\alpha_i),1) \zeta _{0,-1}(m_c)-24 m_c^2 q^2 u_0 i_1(\alpha_g \tilde{\psi}(\alpha_i),v)
\zeta _{0,-1}(m_c)\nonumber\\
&-&24 m_c^2 q^2 u_0 i_1(\alpha_q \tilde{\psi}(\alpha_i),1) \zeta _{0,-1}(m_c)+4 m_c^4
j_1({\cal A}^\parallel (u)) \zeta _{0,-1}(m_c)+m_c^2 M^2 u_0 {\cal A}^\parallel(u_0) \zeta _{0,0}(m_c)\nonumber\\
&+&20 m_c^2 M^2
u_0 \tilde{j}_1({\cal C}(u)) \zeta _{0,0}(m_c)-4 m_c^2 M^2 i_1(\tilde{\phi}(\alpha_i),1) \zeta _{0,0}(m_c)+4
M^4 i_1(\tilde{\phi}(\alpha_i),1) \zeta _{0,0}(m_c)\nonumber\\
&+&4 m_c^2 M^2 i_1(\psi(\alpha_i),1) \zeta _{0,0}(m_c)-12
M^4 i_1(\psi(\alpha_i),1) \zeta _{0,0}(m_c)-8 m_c^2 q^2 u_0^2 i_1(\psi(\alpha_i),1) \zeta _{0,0}(m_c)\nonumber\\
&+&16
M^2 q^2 u_0^2 i_1(\psi(\alpha_i),1) \zeta _{0,0}(m_c)-8 m_c^2 M^2 i_1(\psi(\alpha_i),v)
\zeta _{0,0}(m_c)+24 M^4 i_1(\psi(\alpha_i),v) \zeta _{0,0}(m_c)\nonumber\\
&+&16 m_c^2 q^2 u_0^2 i_1(\psi(\alpha_i),v)
\zeta _{0,0}(m_c)-32 M^2 q^2 u_0^2 i_1(\psi(\alpha_i),v) \zeta _{0,0}(m_c)\nonumber\\
&+&8 m_c^2 q^2 u_0
i_1(\alpha_g \psi(\alpha_i),v) \zeta _{0,0}(m_c)+12 M^4
i_1(\tilde{\psi}(\alpha_i),1) \zeta _{0,0}(m_c)\nonumber\\
&-&16 M^2 q^2 u_0 i_1(\alpha_g \psi(\alpha_i),v) \zeta
_{0,0}(m_c)-16 m_c^2 q^2 u_0 i_1(\alpha_g \psi(\alpha_i),v^2) \zeta _{0,0}(m_c)\nonumber\\
&+&32 M^2
q^2 u_0 i_1(\alpha_g \psi(\alpha_i),v^2) \zeta _{0,0}(m_c)+8 m_c^2 q^2 u_0 i_1(\alpha_q
\psi(\alpha_i),1) \zeta _{0,0}(m_c)\nonumber\\
&-&16 M^2 q^2 u_0 i_1(\alpha_q \psi(\alpha_i),1) \zeta _{0,0}(m_c)-16
m_c^2 q^2 u_0 i_1(\alpha_q \psi(\alpha_i),v) \zeta _{0,0}(m_c)\nonumber\\
&+&32 M^2 q^2 u_0 i_1(\alpha_q
\psi(\alpha_i),v) \zeta _{0,0}(m_c)-4 m_c^2 M^2 i_1(\tilde{\psi}(\alpha_i),1) \zeta _{0,0}(m_c)\nonumber\\
&-&24 m_c^2 q^2 u_0^2 i_1(\tilde{\psi}(\alpha_i),1) \zeta _{0,0}(m_c)+48
M^2 q^2 u_0^2 i_1(\tilde{\psi}(\alpha_i),1) \zeta _{0,0}(m_c)
\nonumber\\
&+&24 m_c^2 q^2 u_0 i_1(\alpha_g
\tilde{\psi}(\alpha_i),v) \zeta _{0,0}(m_c)
-48 M^2 q^2 u_0 i_1(\alpha_g \tilde{\psi}(\alpha_i),v) \zeta _{0,0}(m_c)
\nonumber\\
&+&24m_c^2 q^2 u_0 i_1(\alpha_q \tilde{\psi}(\alpha_i),1) \zeta _{0,0}(m_c)
-48 M^2 q^2 u_0 i_1(\alpha_q
\tilde{\psi}(\alpha_i),1) \zeta _{0,0}(m_c)\nonumber\\
&+&4 M^4 i_2(\alpha_g \phi(\alpha_i),v) \zeta _{0,0}(m_c)-8 M^4
i_2(\alpha_g \phi(\alpha_i),v^2) \zeta _{0,0}(m_c)
\nonumber\\
&+&4 M^4 i_2(\alpha_q \phi(\alpha_i),1) \zeta _{0,0}(m_c)-8
M^4 i_2(\alpha_q \phi(\alpha_i),v) \zeta _{0,0}(m_c)+4 M^4 i_2(\alpha_g \tilde{\phi}(\alpha_i),v) \zeta
_{0,0}(m_c)\nonumber\\
&+&4 M^4 i_2(\alpha_q \tilde{\phi}(\alpha_i),1) \zeta _{0,0}(m_c)+4 M^4 i_2(\alpha_g \psi(\alpha_i),v)
\zeta _{0,0}(m_c)-8 M^4 i_2(\alpha_g \psi(\alpha_i),v^2) \zeta _{0,0}(m_c)\nonumber\\
&+&4 M^4 i_2(\alpha_q
\psi(\alpha_i),1) \zeta _{0,0}(m_c)-8 M^4 i_2(\alpha_q \psi(\alpha_i),v) \zeta _{0,0}(m_c)+4 M^4
i_2(\alpha_g \tilde{\psi}(\alpha_i),v) \zeta _{0,0}(m_c)\nonumber\\
&+&4 M^4 i_2(\alpha_q \tilde{\psi}(\alpha_i),1) \zeta _{0,0}(m_c)-m_c^4
j_1({\cal A}^\parallel (u)) \zeta _{0,0}(m_c)+5 m_c^2 M^2 j_1({\cal A}^\parallel (u)) \zeta _{0,0}(m_c)\nonumber\\
&+&m_c^4 j_1({\cal A}^\parallel (u)) \zeta
_{1,-2}(m_c)+m_c^2 M^2 u_0 {\cal A}^\parallel(u_0) \zeta _{1,-1}(m_c)+12 m_c^2 M^2 u_0 \tilde{j}_1({\cal C}(u))
\zeta _{1,-1}(m_c)\nonumber\\
&-&4 m_c^2 M^2 i_1(\tilde{\phi}(\alpha_i),1) \zeta _{1,-1}(m_c)+4 m_c^2 M^2 i_1(\psi(\alpha_i),1)
\zeta _{1,-1}(m_c)-8 m_c^2 q^2 u_0^2 i_1(\psi(\alpha_i),1) \zeta _{1,-1}(m_c)\nonumber\\
&-&8 m_c^2 M^2
i_1(\psi(\alpha_i),v) \zeta _{1,-1}(m_c)+16 m_c^2 q^2 u_0^2 i_1(\psi(\alpha_i),v) \zeta _{1,-1}(m_c)\nonumber\\
&+&8
m_c^2 q^2 u_0 i_1(\alpha_g \psi(\alpha_i),v) \zeta _{1,-1}(m_c)-16 m_c^2 q^2 u_0 i_1(\alpha_g
\psi(\alpha_i),v^2) \zeta _{1,-1}(m_c)\nonumber\\
&+&8 m_c^2 q^2 u_0 i_1(\alpha_q \psi(\alpha_i),1) \zeta _{1,-1}(m_c)-16
m_c^2 q^2 u_0 i_1(\alpha_q \psi(\alpha_i),v) \zeta _{1,-1}(m_c)\nonumber\\
&-&4 m_c^2 M^2 i_1(\tilde{\psi}(\alpha_i),1)
\zeta _{1,-1}(m_c)-24 m_c^2 q^2 u_0^2 i_1(\tilde{\psi}(\alpha_i),1) \zeta _{1,-1}(m_c)\nonumber\\
&+&24 m_c^2 q^2 u_0 i_1(\alpha_q \tilde{\psi}(\alpha_i),1)
\zeta _{1,-1}(m_c)-2 m_c^4 j_1({\cal A}^\parallel (u)) \zeta _{1,-1}(m_c)+3 m_c^2 M^2 j_1({\cal A}^\parallel (u)) \zeta _{1,-1}(m_c)\nonumber\\
&-&m_c^2
M^2 u_0 {\cal A}^\parallel(u_0) \zeta _{1,0}(m_c)+3 M^4 u_0 {\cal A}^\parallel(u_0) \zeta _{1,0}(m_c)-12 m_c^2
M^2 u_0 \tilde{j}_1({\cal C}(u)) \zeta _{1,0}(m_c)\nonumber\\
&+&48 M^4 u_0 \tilde{j}_1({\cal C}(u)) \zeta _{1,0}(m_c)-4
M^4 i_1(\tilde{\phi}(\alpha_i),1) \zeta _{1,0}(m_c)+12 M^4 i_1(\psi(\alpha_i),1) \zeta _{1,0}(m_c)\nonumber\\
&-&16
M^2 q^2 u_0^2 i_1(\psi(\alpha_i),1) \zeta _{1,0}(m_c)-24 M^4 i_1(\psi(\alpha_i),v) \zeta _{1,0}(m_c)+32
M^2 q^2 u_0^2 i_1(\psi(\alpha_i),v) \zeta _{1,0}(m_c)\nonumber\\
&+&16 M^2 q^2 u_0 i_1(\alpha_g \psi(\alpha_i),v)
\zeta _{1,0}(m_c)+24 m_c^2 q^2
u_0 i_1(\alpha_g \tilde{\psi}(\alpha_i),v) \zeta _{1,-1}(m_c)\nonumber\\
&-&32 M^2 q^2 u_0 i_1(\alpha_g \psi(\alpha_i),v^2) \zeta _{1,0}(m_c)+16 M^2
q^2 u_0 i_1(\alpha_q \psi(\alpha_i),1) \zeta _{1,0}(m_c)\nonumber\\
&-&32 M^2 q^2 u_0 i_1(\alpha_q \psi(\alpha_i),v)
\zeta _{1,0}(m_c)-12 M^4 i_1(\tilde{\psi}(\alpha_i),1) \zeta _{1,0}(m_c)-48 M^2 q^2 u_0^2 i_1(\tilde{\psi}(\alpha_i),1)
\zeta _{1,0}(m_c)\nonumber\\
&+&48 M^2 q^2 u_0 i_1(\alpha_g \tilde{\psi}(\alpha_i),v) \zeta _{1,0}(m_c)+48 M^2
q^2 u_0 i_1(\alpha_q \tilde{\psi}(\alpha_i),1) \zeta _{1,0}(m_c)\nonumber\\
&-&4 M^4 i_2(\alpha_g \phi(\alpha_i),v)
\zeta _{1,0}(m_c)+8 M^4 i_2(\alpha_g \phi(\alpha_i),v^2) \zeta _{1,0}(m_c)\nonumber\\
&-&4 M^4 i_2(\alpha_q
\phi(\alpha_i),1) \zeta _{1,0}(m_c)+8 M^4 i_2(\alpha_q \phi(\alpha_i),v) \zeta _{1,0}(m_c)\nonumber\\
&-&4 M^4
i_2(\alpha_g \tilde{\phi}(\alpha_i),v) \zeta _{1,0}(m_c)-4 M^4 i_2(\alpha_q \tilde{\phi}(\alpha_i),1) \zeta _{1,0}(m_c)-4
M^4 i_2(\alpha_g \psi(\alpha_i),v) \zeta _{1,0}(m_c)\nonumber\\
&+&8 M^4 i_2(\alpha_g \psi(\alpha_i),v^2)
\zeta _{1,0}(m_c)-4 M^4 i_2(\alpha_q \psi(\alpha_i),1) \zeta _{1,0}(m_c)+8 M^4 i_2(\alpha_q \psi(\alpha_i),v)
\zeta _{1,0}(m_c)\nonumber\\
&-&4 M^4 i_2(\alpha_g \tilde{\psi}(\alpha_i),v) \zeta _{1,0}(m_c)-4 M^4 i_2(\alpha_q \tilde{\psi}(\alpha_i),1)
\zeta _{1,0}(m_c)-3 m_c^2 M^2 j_1({\cal A}^\parallel (u)) \zeta _{1,0}(m_c)\nonumber\\
&+&3 M^4 j_1({\cal A}^\parallel (u)) \zeta _{1,0}(m_c)+8
M^2 i_1(\phi(\alpha_i),v) (m_c^2 \zeta _{-1,0}(m_c)+m_c^2 \zeta _{0,-1}(m_c)-m_c^2 \zeta _{0,0}(m_c)\nonumber\\
&+&M^2
\zeta _{0,0}(m_c)-m_c^2 \zeta _{1,-1}(m_c)-M^2 \zeta _{1,0}(m_c))+4 M^2 i_1(\phi(\alpha_i),1)
(-m_c^2 \zeta _{-1,0}(m_c)-m_c^2 \zeta _{0,-1}(m_c)\nonumber\\
&+&m_c^2 \zeta _{0,0}(m_c)-M^2 \zeta _{0,0}(m_c)+m_c^2
\zeta _{1,-1}(m_c)+M^2 \zeta _{1,0}(m_c))-m_c^4 j_1({\cal A}^\parallel (u)) \zeta _{2,-2}(m_c)\nonumber\\
&-&m_c^2 M^2
u_0 {\cal A}^\parallel(u_0) \zeta _{2,-1}(m_c)-12 m_c^2 M^2 u_0 \tilde{j}_1({\cal C}(u)) \zeta _{2,-1}(m_c)-3
m_c^2 M^2 j_1({\cal A}^\parallel (u)) \zeta _{2,-1}(m_c)\nonumber\\
&-&3 M^4 (u_0 {\cal A}^\parallel(u_0)+16 u_0 \tilde{j}_1({\cal C}(u))+j_1({\cal A}^\parallel (u)))
\zeta _{2,0}(m_c)\Bigg)+M^4 \Bigg(-7 q^2 i_2(\alpha_q \tilde{{\cal A}}(\alpha_i),1) \zeta _{0,0}(m_c)\nonumber\\
&+&8 q^2 i_2(\alpha_q
\tilde{{\cal A}}(\alpha_i),v) \zeta _{0,0}(m_c)+10 q^2 i_2(\alpha_g {\cal V}(\alpha_i),v) \zeta _{0,0}(m_c)-12 q^2 i_2(\alpha_g
{\cal V}(\alpha_i),v^2) \zeta _{0,0}(m_c)\nonumber\\
&+&10 q^2 i_2(\alpha_q {\cal V}(\alpha_i),1) \zeta _{0,0}(m_c)-12 q^2
i_2(\alpha_q {\cal V}(\alpha_i),v) \zeta _{0,0}(m_c)-3 M^2 i_3(\tilde{{\cal A}}(\alpha_i),1) \zeta _{0,0}(m_c)\nonumber\\
&+&3 M^2
i_3(\tilde{{\cal A}}(\alpha_i),v) \zeta _{0,0}(m_c)+2 M^2 i_3({\cal V}(\alpha_i),1) \zeta _{0,0}(m_c)-2 M^2 i_3({\cal V}(\alpha_i),v)
\zeta _{0,0}(m_c)\nonumber\\
&+&12 m_c^2 j_1({\cal B}(u)) \zeta _{0,0}(m_c)-12 m_c^2 j_1(\phi^\parallel (u)) \zeta _{0,0}(m_c)+4
m_c^2 j_1({\cal B}(u)) \zeta _{1,-1}(m_c)\nonumber\\
&-&4 m_c^2 j_1(\phi^\parallel (u)) \zeta _{1,-1}(m_c)+8 q^2 i_2(\alpha_g
\tilde{{\cal A}}(\alpha_i),v^2) (\zeta _{0,0}(m_c)-\zeta _{1,0}(m_c))-4 M^2 u_0 \phi^\parallel(u_0)
\zeta _{1,0}(m_c)\nonumber\\
&-&2 M^2 \psi_3^\perp(u_0) \zeta _{1,0}(m_c)+7 q^2 i_2(\alpha_q \tilde{{\cal A}}(\alpha_i),1) \zeta
_{1,0}(m_c)-8 q^2 i_2(\alpha_q \tilde{{\cal A}}(\alpha_i),v) \zeta _{1,0}(m_c)\nonumber\\
&-&10 q^2 i_2(\alpha_g {\cal V}(\alpha_i),v)
\zeta _{1,0}(m_c)+12 q^2 i_2(\alpha_g {\cal V}(\alpha_i),v^2) \zeta _{1,0}(m_c)-10 q^2 i_2(\alpha_q
{\cal V}(\alpha_i),1) \zeta _{1,0}(m_c)\nonumber\\
&+&12 q^2 i_2(\alpha_q {\cal V}(\alpha_i),v) \zeta _{1,0}(m_c)+3 M^2 i_3(\tilde{{\cal A}}(\alpha_i),1)
\zeta _{1,0}(m_c)-3 M^2 i_3(\tilde{{\cal A}}(\alpha_i),v) \zeta _{1,0}(m_c)\nonumber\\
&-&2 M^2 i_3({\cal V}(\alpha_i),1) \zeta _{1,0}(m_c)+2
M^2 i_3({\cal V}(\alpha_i),v) \zeta _{1,0}(m_c)-4 m_c^2 j_1({\cal B}(u)) \zeta _{1,0}(m_c)\nonumber\\
&+&8 M^2 j_1({\cal B}(u))
\zeta _{1,0}(m_c)+4 m_c^2 j_1(\phi^\parallel (u)) \zeta _{1,0}(m_c)-8 M^2 j_1(\phi^\parallel (u)) \zeta _{1,0}(m_c)\nonumber\\
&+&7
q^2 i_2(\alpha_g \tilde{{\cal A}}(\alpha_i),v) (-\zeta _{0,0}(m_c)+\zeta _{1,0}(m_c))-4 m_c^2 j_1({\cal B}(u))
\zeta _{2,-1}(m_c)\nonumber\\
&+&4 m_c^2 j_1(\phi^\parallel (u)) \zeta _{2,-1}(m_c)+2 M^2 \Big[2 u_0 \phi^\parallel(u_0)+\psi_3^\perp(u_0)\nonumber\\
&-&4
j_1({\cal B}(u))+4 j_1(\phi^\parallel (u))\Big] \zeta _{2,0}(m_c)\Bigg)\Bigg]\Biggr\}~,
\end{eqnarray}

\begin{eqnarray}
\Pi_A^{GG}&=&-\frac{ e^{-\frac{q^2}{M_1^2+M_2^2}}}{1728 m_c M^8 \pi ^2} g_s^2\langle GG\rangle \Biggl\{2 m_c f_\rho^\parallel m_\rho \Bigg[2 m_\rho^2
\Biggl(-8 m_c^4 q^2 u_0 i_1(\alpha_g \tilde{\psi}(\alpha_i),v) \zeta _{-3,0}(m_c)\nonumber\\
&-&8 m_c^4 q^2 u_0
i_1(\alpha_q \tilde{\psi}(\alpha_i),1) \zeta _{-3,0}(m_c)
+3 m_c^6 j_1({\cal A}^\parallel(u)) \zeta _{-3,0}(m_c)\nonumber\\
&-&16 m_c^4
q^2 u_0 i_1(\alpha_g \tilde{\psi}(\alpha_i),v) \zeta _{-2,-1}(m_c)-16 m_c^4 q^2 u_0 i_1(\alpha_q
\tilde{\psi}(\alpha_i),1) \zeta _{-2,-1}(m_c)\nonumber\\
&+&9 m_c^6 j_1({\cal A}^\parallel(u)) \zeta _{-2,-1}(m_c)+m_c^4 M^2 u_0
{\cal A}^\parallel(u_0) \zeta _{-2,0}(m_c)\nonumber\\
&+&20 m_c^4 M^2 u_0 \tilde{j}_1({\cal C}(u)) \zeta _{-2,0}(m_c)-2 m_c^4
M^2 i_1(\phi(\alpha_i),1) \zeta _{-2,0}(m_c)\nonumber\\
&+&4 m_c^4 M^2 i_1(\phi(\alpha_i),v) \zeta _{-2,0}(m_c)-2
m_c^4 M^2 i_1(\psi(\alpha_i),1) \zeta _{-2,0}(m_c)\nonumber\\
&+&4 m_c^4 q^2 u_0^2 i_1(\psi(\alpha_i),1)
\zeta _{-2,0}(m_c)+4 m_c^4 M^2 i_1(\psi(\alpha_i),v) \zeta _{-2,0}(m_c)\nonumber\\
&-&8 m_c^4 q^2 u_0^2
i_1(\psi(\alpha_i),v) \zeta _{-2,0}(m_c)-4 m_c^4 q^2 u_0 i_1(\alpha_g \psi(\alpha_i),v) \zeta _{-2,0}(m_c)\nonumber\\
&+&8
m_c^4 q^2 u_0 i_1(\alpha_g \psi(\alpha_i),v^2) \zeta _{-2,0}(m_c)-4 m_c^4 q^2 u_0
i_1(\alpha_q \psi(\alpha_i),1) \zeta _{-2,0}(m_c)\nonumber\\
&+&8 m_c^4 q^2 u_0 i_1(\alpha_q \psi(\alpha_i),v) \zeta
_{-2,0}(m_c)-4 m_c^4 q^2 u_0 i_1(\alpha_g \tilde{\psi}(\alpha_i),v) \zeta _{-2,0}(m_c)\nonumber\\
&+&24 m_c^2 M^2
q^2 u_0 i_1(\alpha_g \tilde{\psi}(\alpha_i),v) \zeta _{-2,0}(m_c)-4 m_c^4 q^2 u_0 i_1(\alpha_q \tilde{\psi}(\alpha_i),1)
\zeta _{-2,0}(m_c)\nonumber\\
&+&24 m_c^2 M^2 q^2 u_0 i_1(\alpha_q \tilde{\psi}(\alpha_i),1) \zeta _{-2,0}(m_c)-3
m_c^6 j_1({\cal A}^\parallel(u)) \zeta _{-2,0}(m_c)\nonumber\\
&-&7 m_c^4 M^2 j_1({\cal A}^\parallel(u)) \zeta _{-2,0}(m_c)-8 m_c^4 q^2
u_0 i_1(\alpha_g \tilde{\psi}(\alpha_i),v) \zeta _{-1,-2}(m_c)\nonumber\\
&-&8 m_c^4 q^2 u_0 i_1(\alpha_q \tilde{\psi}(\alpha_i),1)
\zeta _{-1,-2}(m_c)+9 m_c^6 j_1({\cal A}^\parallel(u)) \zeta _{-1,-2}(m_c)\nonumber\\
&+&2 m_c^4 M^2 u_0 {\cal A}^\parallel(u_0) \zeta
_{-1,-1}(m_c)+40 m_c^4 M^2 u_0 \tilde{j}_1({\cal C}(u)) \zeta _{-1,-1}(m_c)\nonumber\\
&-&4 m_c^4 M^2 i_1(\phi(\alpha_i),1)
\zeta _{-1,-1}(m_c)+8 m_c^4 M^2 i_1(\phi(\alpha_i),v) \zeta _{-1,-1}(m_c)\nonumber\\
&-&4 m_c^4 M^2 i_1(\psi(\alpha_i),1)
\zeta _{-1,-1}(m_c)+8 m_c^4 q^2 u_0^2 i_1(\psi(\alpha_i),1) \zeta _{-1,-1}(m_c)\nonumber\\
&+&8 m_c^4 M^2
i_1(\psi(\alpha_i),v) \zeta _{-1,-1}(m_c)-16 m_c^4 q^2 u_0^2 i_1(\psi(\alpha_i),v) \zeta _{-1,-1}(m_c)\nonumber\\
&-&8
m_c^4 q^2 u_0 i_1(\alpha_g \psi(\alpha_i),v) \zeta _{-1,-1}(m_c)+16 m_c^4 q^2 u_0 i_1(\alpha_g
\psi(\alpha_i),v^2) \zeta _{-1,-1}(m_c)\nonumber\\
&-&8 m_c^4 q^2 u_0 i_1(\alpha_q \psi(\alpha_i),1) \zeta _{-1,-1}(m_c)+16
m_c^4 q^2 u_0 i_1(\alpha_q \psi(\alpha_i),v) \zeta _{-1,-1}(m_c)\nonumber\\
&-&8 m_c^4 q^2 u_0 i_1(\alpha_g
\tilde{\psi}(\alpha_i),v) \zeta _{-1,-1}(m_c)+24 m_c^2 M^2 q^2 u_0 i_1(\alpha_g \tilde{\psi}(\alpha_i),v)
\zeta _{-1,-1}(m_c)\nonumber\\
&-&8 m_c^4 q^2 u_0 i_1(\alpha_q \tilde{\psi}(\alpha_i),1) \zeta _{-1,-1}(m_c)+24 m_c^2
M^2 q^2 u_0 i_1(\alpha_q \tilde{\psi}(\alpha_i),1) \zeta _{-1,-1}(m_c)\nonumber\\
&-&9 m_c^6 j_1({\cal A}^\parallel(u)) \zeta _{-1,-1}(m_c)-14
m_c^4 M^2 j_1({\cal A}^\parallel(u)) \zeta _{-1,-1}(m_c)-m_c^4 M^2 u_0 {\cal A}^\parallel(u_0) \zeta _{-1,0}(m_c)\nonumber\\
&+&2
m_c^2 M^4 u_0 {\cal A}^\parallel(u_0) \zeta _{-1,0}(m_c)-20 m_c^4 M^2 u_0 \tilde{j}_1({\cal C}(u))
\zeta _{-1,0}(m_c)\nonumber\\
&+&28 m_c^2 M^4 u_0 \tilde{j}_1({\cal C}(u)) \zeta _{-1,0}(m_c)+2 m_c^2 M^4 i_1(\phi(\alpha_i),1)
\zeta _{-1,0}(m_c)\nonumber\\
&-&4 m_c^2 M^4 i_1(\phi(\alpha_i),v) \zeta _{-1,0}(m_c)-2 m_c^2 M^4 i_1(\psi(\alpha_i),1)
\zeta _{-1,0}(m_c)\nonumber\\
&+&4 m_c^2 M^4 i_1(\psi(\alpha_i),v) \zeta _{-1,0}(m_c)
+2 m_c^2 M^4 i_2(\alpha_g
\phi(\alpha_i),v) \zeta _{-1,0}(m_c)\nonumber\\
&-&4 m_c^2 M^4 i_2(\alpha_g \phi(\alpha_i),v^2) \zeta _{-1,0}(m_c)
+2
m_c^2 M^4 i_2(\alpha_q \phi(\alpha_i),1) \zeta _{-1,0}(m_c)\nonumber\\
&-&4 m_c^2 M^4 i_2(\alpha_q \phi(\alpha_i),v)
\zeta _{-1,0}(m_c)+2 m_c^2 M^4 i_2(\alpha_g \tilde{\phi}(\alpha_i),v) \zeta _{-1,0}(m_c)\nonumber\\
&+&2 m_c^2 M^4
i_2(\alpha_q \tilde{\phi}(\alpha_i),1) \zeta _{-1,0}(m_c)+2 m_c^2 M^4 i_2(\alpha_g \psi(\alpha_i),v) \zeta
_{-1,0}(m_c)\nonumber\\
&-&4 m_c^2 M^4 i_2(\alpha_g \psi(\alpha_i),v^2) \zeta _{-1,0}(m_c)+2 m_c^2 M^4
i_2(\alpha_q \psi(\alpha_i),1) \zeta _{-1,0}(m_c)\nonumber\\
&-&4 m_c^2 M^4 i_2(\alpha_q \psi(\alpha_i),v) \zeta _{-1,0}(m_c)+2
m_c^2 M^4 i_2(\alpha_g \tilde{\psi}(\alpha_i),v) \zeta _{-1,0}(m_c)\nonumber\\
&+&2 m_c^2 M^4 i_2(\alpha_q \tilde{\psi}(\alpha_i),1)
\zeta _{-1,0}(m_c)+m_c^4 M^2 j_1({\cal A}^\parallel(u)) \zeta _{-1,0}(m_c)\nonumber\\
&-&2 m_c^2 M^4 j_1({\cal A}^\parallel(u)) \zeta
_{-1,0}(m_c)+3 m_c^6 j_1({\cal A}^\parallel(u)) \zeta _{0,-3}(m_c)+m_c^4 M^2 u_0 {\cal A}^\parallel(u_0) \zeta _{0,-2}(m_c)\nonumber\\
&+&20
m_c^4 M^2 u_0 \tilde{j}_1({\cal C}(u)) \zeta _{0,-2}(m_c)-2 m_c^4 M^2 i_1(\phi(\alpha_i),1) \zeta
_{0,-2}(m_c)\nonumber\\
&+&4 m_c^4 M^2 i_1(\phi(\alpha_i),v) \zeta _{0,-2}(m_c)-2 m_c^4 M^2 i_1(\psi(\alpha_i),1)
\zeta _{0,-2}(m_c)\nonumber\\
&+&4 m_c^4 M^2
i_1(\psi(\alpha_i),v) \zeta _{0,-2}(m_c)-8 m_c^4 q^2 u_0^2 i_1(\psi(\alpha_i),v) \zeta _{0,-2}(m_c)\nonumber\\
&-&4m_c^4 q^2 u_0 i_1(\alpha_g \psi(\alpha_i),v) \zeta _{0,-2}(m_c)
+8 m_c^4 q^2 u_0 i_1(\alpha_g
\psi(\alpha_i),v^2) \zeta _{0,-2}(m_c)\nonumber\\
&-&4 m_c^4 q^2 u_0 i_1(\alpha_q \psi(\alpha_i),1) \zeta _{0,-2}(m_c)
+8
m_c^4 q^2 u_0 i_1(\alpha_q \psi(\alpha_i),v) \zeta _{0,-2}(m_c)\nonumber\\
&-&4 m_c^4 q^2 u_0 i_1(\alpha_g
\tilde{\psi}(\alpha_i),v) \zeta _{0,-2}(m_c)-4 m_c^4 q^2 u_0 i_1(\alpha_g
\tilde{\psi}(\alpha_i),v) \zeta _{0,-2}(m_c)\nonumber\\
&-&4 m_c^4 q^2 u_0 i_1(\alpha_q \tilde{\psi}(\alpha_i),1) \zeta
_{0,-2}(m_c)-9 m_c^6 j_1({\cal A}^\parallel(u)) \zeta _{0,-2}(m_c)\nonumber\\
&-&7 m_c^4 M^2 j_1({\cal A}^\parallel(u)) \zeta _{0,-2}(m_c)-2
m_c^4 M^2 u_0 {\cal A}^\parallel(u_0) \zeta _{0,-1}(m_c)\nonumber\\
&+&2 m_c^2 M^4 u_0 {\cal A}^\parallel(u_0) \zeta
_{0,-1}(m_c)-40 m_c^4 M^2 u_0 \tilde{j}_1({\cal C}(u)) \zeta _{0,-1}(m_c)\nonumber\\
&+&28 m_c^2 M^4 u_0
\tilde{j}_1({\cal C}(u)) \zeta _{0,-1}(m_c)+2 m_c^2 M^4 i_1(\phi(\alpha_i),1) \zeta _{0,-1}(m_c)\nonumber\\
&+&4 m_c^2
M^4 i_1(\phi(\alpha_i),v) \zeta _{0,-1}(m_c)-2 m_c^2 M^4 i_1(\psi(\alpha_i),1) \zeta _{0,-1}(m_c)\nonumber\\
&+&
m_c^2 M^4 i_1(\psi(\alpha_i),v) \zeta _{0,-1}(m_c)+2 m_c^2 M^4 i_2(\alpha_g \phi(\alpha_i),v)
\zeta _{0,-1}(m_c)\nonumber\\
&-&4 m_c^2 M^4 i_2(\alpha_g \phi(\alpha_i),v^2) \zeta _{0,-1}(m_c)+2 m_c^2
M^4 i_2(\alpha_q \phi(\alpha_i),1) \zeta _{0,-1}(m_c)\nonumber\\
&-&4 m_c^2 M^4 i_2(\alpha_q \phi(\alpha_i),v)
\zeta _{0,-1}(m_c)+2 m_c^2 M^4 i_2(\alpha_g \tilde{\phi}(\alpha_i),v) \zeta _{0,-1}(m_c)\nonumber\\
&+&2 m_c^2 M^4
i_2(\alpha_q \tilde{\phi}(\alpha_i),1) \zeta _{0,-1}(m_c)+2 m_c^2 M^4 i_2(\alpha_g \psi(\alpha_i),v) \zeta
_{0,-1}(m_c)\nonumber\\
&-&4 m_c^2 M^4 i_2(\alpha_g \psi(\alpha_i),v^2) \zeta _{0,-1}(m_c)+2 m_c^2 M^4
i_2(\alpha_q \psi(\alpha_i),1) \zeta _{0,-1}(m_c)\nonumber\\
&-&4 m_c^2 M^4 i_2(\alpha_q \psi(\alpha_i),v) \zeta _{0,-1}(m_c)+4 m_c^4 q^2 u_0^2 i_1(\psi(\alpha_i),1) \zeta _{0,-2}(m_c)\nonumber\\
&+&2
m_c^2 M^4 i_2(\alpha_g \tilde{\psi}(\alpha_i),v) \zeta _{0,-1}(m_c)+2 m_c^2 M^4 i_2(\alpha_q \tilde{\psi}(\alpha_i),1)
\zeta _{0,-1}(m_c)\nonumber\\
&+&2 m_c^4 M^2 j_1({\cal A}^\parallel(u)) \zeta _{0,-1}(m_c)-2 m_c^2 M^4 j_1({\cal A}^\parallel(u)) \zeta
_{0,-1}(m_c)-2 m_c^2 M^4 u_0 {\cal A}^\parallel(u_0) \zeta _{0,0}(m_c)\nonumber\\
&+&2 M^6 u_0 {\cal A}^\parallel(u_0)
\zeta _{0,0}(m_c)-52 m_c^2 M^4 u_0 \tilde{j}_1({\cal C}(u)) \zeta _{0,0}(m_c)+40 M^6 u_0 \tilde{j}_1({\cal C}(u))
\zeta _{0,0}(m_c)\nonumber\\
&+&2 M^6 i_1(\phi(\alpha_i),1) \zeta _{0,0}(m_c)-4 M^6 i_1(\phi(\alpha_i),v) \zeta
_{0,0}(m_c)-2 M^6 i_1(\psi(\alpha_i),1) \zeta _{0,0}(m_c)\nonumber\\
&+&4 M^6 i_1(\psi(\alpha_i),v) \zeta _{0,0}(m_c)+2
M^6 i_2(\alpha_g \phi(\alpha_i),v) \zeta _{0,0}(m_c)-4 M^6 i_2(\alpha_g \phi(\alpha_i),v^2)
\zeta _{0,0}(m_c)\nonumber\\
&+&2 M^6 i_2(\alpha_q \phi(\alpha_i),1) \zeta _{0,0}(m_c)-4 M^6 i_2(\alpha_q \phi(\alpha_i),v)
\zeta _{0,0}(m_c)+2 M^6 i_2(\alpha_g \tilde{\phi}(\alpha_i),v) \zeta _{0,0}(m_c)\nonumber\\
&+&2 M^6 i_2(\alpha_q \tilde{\phi}(\alpha_i),1)
\zeta _{0,0}(m_c)+2 M^6 i_2(\alpha_g \psi(\alpha_i),v) \zeta _{0,0}(m_c)-4 M^6 i_2(\alpha_g \psi(\alpha_i),v^2)
\zeta _{0,0}(m_c)\nonumber\\
&+&2 M^6 i_2(\alpha_q \psi(\alpha_i),1) \zeta _{0,0}(m_c)-4 M^6 i_2(\alpha_q \psi(\alpha_i),v)
\zeta _{0,0}(m_c)+2 M^6 i_2(\alpha_g \tilde{\psi}(\alpha_i),v) \zeta _{0,0}(m_c)\nonumber\\
&+&2 M^6 i_2(\alpha_q \tilde{\psi}(\alpha_i),1)
\zeta _{0,0}(m_c)+2 m_c^2 M^4 j_1({\cal A}^\parallel(u)) \zeta _{0,0}(m_c)-2 M^6 j_1({\cal A}^\parallel(u)) \zeta _{0,0}(m_c)\nonumber\\
&-&2
M^2 i_1(\tilde{\phi}(\alpha_i),1) (-2 m_c^4 \zeta _{-3,0}(m_c)-4 m_c^4 \zeta _{-2,-1}(m_c)+m_c^2
((m_c^2+6 M^2) \zeta _{-2,0}(m_c)\nonumber\\
&-&2 m_c^2 \zeta _{-1,-2}(m_c)+2 m_c^2 \zeta _{-1,-1}(m_c)+6
M^2 \zeta _{-1,-1}(m_c)+M^2 \zeta _{-1,0}(m_c)+m_c^2 \zeta _{0,-2}(m_c)\nonumber\\
&+&M^2 \zeta _{0,-1}(m_c))+M^4
\zeta _{0,0}(m_c))+2 i_1(\tilde{\psi}(\alpha_i),1) (2 m_c^4 (M^2+2 q^2 u_0^2) \zeta
_{-3,0}(m_c)\nonumber\\
&+&4 m_c^4 (M^2+2 q^2 u_0^2) \zeta _{-2,-1}(m_c)+m_c^2 (-(m_c^2
(M^2-2 q^2 u_0^2)+6 M^2 (M^2+2 q^2 u_0^2)) \zeta _{-2,0}(m_c)\nonumber\\
&+&2
m_c^2 (M^2+2 q^2 u_0^2) \zeta _{-1,-2}(m_c)-2 m_c^2 M^2 \zeta _{-1,-1}(m_c)-6 M^4
\zeta _{-1,-1}(m_c)+4 m_c^2 q^2 u_0^2 \zeta _{-1,-1}(m_c)\nonumber\\
&-&12 M^2 q^2 u_0^2 \zeta _{-1,-1}(m_c)+M^4
\zeta _{-1,0}(m_c)-m_c^2 M^2 \zeta _{0,-2}(m_c)+2 m_c^2 q^2 u_0^2 \zeta _{0,-2}(m_c)\nonumber\\
&+&M^4
\zeta _{0,-1}(m_c))+M^6 \zeta _{0,0}(m_c))-3 m_c^6 j_1({\cal A}^\parallel(u)) \zeta _{1,-3}(m_c)-m_c^4
M^2 u_0 {\cal A}^\parallel(u_0) \zeta _{1,-2}(m_c)\nonumber\\
&-&20 m_c^4 M^2 u_0 \tilde{j}_1({\cal C}(u)) \zeta _{1,-2}(m_c)+m_c^4
M^2 j_1({\cal A}^\parallel(u)) \zeta _{1,-2}(m_c)-2 m_c^2 M^4 u_0 {\cal A}^\parallel(u_0) \zeta _{1,-1}(m_c)\nonumber\\
&-&52 m_c^2
M^4 u_0 \tilde{j}_1({\cal C}(u)) \zeta _{1,-1}(m_c)+2 m_c^2 M^4 j_1({\cal A}^\parallel(u)) \zeta _{1,-1}(m_c)\nonumber\\
&+&2
M^6 (-u_0 ({\cal A}^\parallel(u_0)+32 \tilde{j}_1({\cal C}(u)))+j_1({\cal A}^\parallel(u))) \zeta _{1,0}(m_c)\Biggr) \nonumber\\
&+&M^4
\Biggl(-8 m_c^2 M^2 u_0 \phi^{\parallel}(u_0) \zeta _{-1,0}(m_c)-4 m_c^2 M^2 \psi_3^\perp(u_0)
\zeta _{-1,0}(m_c)\nonumber\\
&-&7 m_c^2 q^2 i_2(\alpha_g \tilde{A}(\alpha_i),v) \zeta _{-1,0}(m_c)+8 m_c^2 q^2
i_2(\alpha_g \tilde{A}(\alpha_i),v^2) \zeta _{-1,0}(m_c)\nonumber\\
&-&7 m_c^2 q^2 i_2(\alpha_q \tilde{A}(\alpha_i),1)
\zeta _{-1,0}(m_c)+8 m_c^2 q^2 i_2(\alpha_q \tilde{A}(\alpha_i),v) \zeta _{-1,0}(m_c)\nonumber\\
&+&10 m_c^2 q^2
i_2(\alpha_g {\cal V}(\alpha_i),v) \zeta _{-1,0}(m_c)-12 m_c^2 q^2 i_2(\alpha_g {\cal V}(\alpha_i),v^2)
\zeta _{-1,0}(m_c)\nonumber\\
&+&10 m_c^2 q^2 i_2(\alpha_q {\cal V}(\alpha_i),1) \zeta _{-1,0}(m_c)-12 m_c^2 q^2
i_2(\alpha_q {\cal V}(\alpha_i),v) \zeta _{-1,0}(m_c)\nonumber\\
&-&3 m_c^2 M^2 i_3(\tilde{A}(\alpha_i),1) \zeta _{-1,0}(m_c)+3
m_c^2 M^2 i_3(\tilde{A}(\alpha_i),v) \zeta _{-1,0}(m_c)\nonumber\\
&+&2 m_c^2 M^2 i_3({\cal V}(\alpha_i),1) \zeta _{-1,0}(m_c)-2
m_c^2 M^2 i_3({\cal V}(\alpha_i),v) \zeta _{-1,0}(m_c)\nonumber\\
&-&8 m_c^2 M^2 u_0 \phi^{\parallel}(u_0) \zeta
_{0,-1}(m_c)-4 m_c^2 M^2 \psi_3^\perp(u_0) \zeta _{0,-1}(m_c)\nonumber\\
&-&7 m_c^2 q^2 i_2(\alpha_g \tilde{A}(\alpha_i),v)
\zeta _{0,-1}(m_c)+8 m_c^2 q^2 i_2(\alpha_g \tilde{A}(\alpha_i),v^2) \zeta _{0,-1}(m_c)\nonumber\\
&-&7 m_c^2
q^2 i_2(\alpha_q \tilde{A}(\alpha_i),1) \zeta _{0,-1}(m_c)+8 m_c^2 q^2 i_2(\alpha_q \tilde{A}(\alpha_i),v)
\zeta _{0,-1}(m_c)\nonumber\\
&+&10 m_c^2 q^2 i_2(\alpha_g {\cal V}(\alpha_i),v) \zeta _{0,-1}(m_c)-12 m_c^2 q^2
i_2(\alpha_g {\cal V}(\alpha_i),v^2) \zeta _{0,-1}(m_c)\nonumber\\
&+&10 m_c^2 q^2 i_2(\alpha_q {\cal V}(\alpha_i),1)
\zeta _{0,-1}(m_c)-12 m_c^2 q^2 i_2(\alpha_q {\cal V}(\alpha_i),v) \zeta _{0,-1}(m_c)\nonumber\\
&-&3 m_c^2 M^2
i_3(\tilde{A}(\alpha_i),1) \zeta _{0,-1}(m_c)+3 m_c^2 M^2 i_3(\tilde{A}(\alpha_i),v) \zeta _{0,-1}(m_c)\nonumber\\
&+&2 m_c^2
M^2 i_3({\cal V}(\alpha_i),1) \zeta _{0,-1}(m_c)-2 m_c^2 M^2 i_3({\cal V}(\alpha_i),v) \zeta _{0,-1}(m_c)\nonumber\\
&+&8
m_c^2 M^2 u_0 \phi^{\parallel}(u_0) \zeta _{0,0}(m_c)-16 M^4 u_0 \phi^{\parallel}(u_0) \zeta _{0,0}(m_c)\nonumber\\
&+&4
m_c^2 M^2 \psi_3^\perp(u_0) \zeta _{0,0}(m_c)-8 M^4 \psi_3^\perp(u_0) \zeta _{0,0}(m_c)-7
M^2 q^2 i_2(\alpha_g \tilde{A}(\alpha_i),v) \zeta _{0,0}(m_c)\nonumber\\
&+&8 M^2 q^2 i_2(\alpha_g \tilde{A}(\alpha_i),v^2)
\zeta _{0,0}(m_c)-7 M^2 q^2 i_2(\alpha_q \tilde{A}(\alpha_i),1) \zeta _{0,0}(m_c)\nonumber\\
&+&8 M^2 q^2 i_2(\alpha_q
\tilde{A}(\alpha_i),v) \zeta _{0,0}(m_c)+10 M^2 q^2 i_2(\alpha_g {\cal V}(\alpha_i),v) \zeta _{0,0}(m_c)\nonumber\\
&-&12 M^2
q^2 i_2(\alpha_g {\cal V}(\alpha_i),v^2) \zeta _{0,0}(m_c)+10 M^2 q^2 i_2(\alpha_q {\cal V}(\alpha_i),1)
\zeta _{0,0}(m_c)\nonumber\\
&-&12 M^2 q^2 i_2(\alpha_q {\cal V}(\alpha_i),v) \zeta _{0,0}(m_c)-3 M^4 i_3(\tilde{A}(\alpha_i),1)
\zeta _{0,0}(m_c)\nonumber\\
&+&3 M^4 i_3(\tilde{A}(\alpha_i),v) \zeta _{0,0}(m_c)+2 M^4 i_3({\cal V}(\alpha_i),1) \zeta
_{0,0}(m_c)\nonumber\\
&-&2 M^4 i_3({\cal V}(\alpha_i),v) \zeta _{0,0}(m_c)+8 m_c^2 M^2 u_0 \phi^{\parallel}(u_0)
\zeta _{1,-1}(m_c)+4 m_c^2 M^2 \psi_3^\perp(u_0) \zeta _{1,-1}(m_c)\nonumber\\
&+&8 M^4 (2 u_0 \phi^{\parallel}(u_0)+\psi_3^\perp(u_0))
\zeta _{1,0}(m_c)+8 j_1({\cal B}(u)) (3 m_c^4 \zeta _{-2,0}(m_c)+6 m_c^4 \zeta _{-1,-1}(m_c)\nonumber\\
&-&3 m_c^4
\zeta _{-1,0}(m_c)-m_c^2 M^2 \zeta _{-1,0}(m_c)+3 m_c^4 \zeta _{0,-2}(m_c)-6 m_c^4 \zeta _{0,-1}(m_c)-m_c^2
M^2 \zeta _{0,-1}(m_c)\nonumber\\
&-&5 m_c^2 M^2 \zeta _{0,0}(m_c)-2 M^4 \zeta _{0,0}(m_c)-3 m_c^4 \zeta
_{1,-2}(m_c)-5 m_c^2 M^2 \zeta _{1,-1}(m_c)-4 M^4 \zeta _{1,0}(m_c))\nonumber\\
&-&8 j_1(\phi^{\parallel}(u)) (3
m_c^4 \zeta _{-2,0}(m_c)+6 m_c^4 \zeta _{-1,-1}(m_c)-3 m_c^4 \zeta _{-1,0}(m_c)-m_c^2 M^2
\zeta _{-1,0}(m_c)\nonumber\\
&+&3 m_c^4 \zeta _{0,-2}(m_c)-6 m_c^4 \zeta _{0,-1}(m_c)-m_c^2 M^2 \zeta _{0,-1}(m_c)-5
m_c^2 M^2 \zeta _{0,0}(m_c)-2 M^4 \zeta _{0,0}(m_c)\nonumber\\
&-&3 m_c^4 \zeta _{1,-2}(m_c)-5 m_c^2
M^2 \zeta _{1,-1}(m_c)-4 M^4 \zeta _{1,0}(m_c))\Biggr)\Bigg]\nonumber\\
&+&M^2 f_\rho^\perp \Bigg[-4 M^4
\phi^\perp(u_0) \Biggl((-2 m_c^4+3 m_c^2 M^2) \zeta _{-1,0}(m_c)+3 m_c^2 (-m_c^2+M^2)
\zeta _{0,-1}(m_c)\nonumber\\
&+&m_c^4 \zeta _{0,0}(m_c)-8 m_c^2 M^2 \zeta _{0,0}(m_c)+6 M^4 \zeta _{0,0}(m_c)-m_c^4
\zeta _{1,-2}(m_c)+2 m_c^4 \zeta _{1,-1}(m_c)\nonumber\\
&-&7 m_c^2 M^2 \zeta _{1,-1}(m_c)+4 m_c^2 M^2
\zeta _{1,0}(m_c)-12 M^4 \zeta _{1,0}(m_c)+m_c^4 \zeta _{2,-2}(m_c)\nonumber\\
&+&4 m_c^2 M^2 \zeta _{2,-1}(m_c)+6
M^4 \zeta _{2,0}(m_c)\Biggr)+{\cal A}^{\perp}(u_0) m_\rho^2 \Bigg((-2 m_c^6+3 m_c^4 M^2) \zeta
_{-2,0}(m_c)\nonumber\\
&+&(-5 m_c^6+6 m_c^4 M^2) \zeta _{-1,-1}(m_c)+m_c^6 \zeta _{-1,0}(m_c)-6 m_c^4
M^2 \zeta _{-1,0}(m_c)+6 m_c^2 M^4 \zeta _{-1,0}(m_c)\nonumber\\
&-&4 m_c^6 \zeta _{0,-2}(m_c)+3 m_c^4
M^2 \zeta _{0,-2}(m_c)+3 m_c^6 \zeta _{0,-1}(m_c)-12 m_c^4 M^2 \zeta _{0,-1}(m_c)\nonumber\\
&+&6 m_c^2
M^4 \zeta _{0,-1}(m_c)+3 m_c^4 M^2 \zeta _{0,0}(m_c)-12 m_c^2 M^4 \zeta _{0,0}(m_c)+6
M^6 \zeta _{0,0}(m_c)\nonumber\\
&-&m_c^6 \zeta _{1,-3}(m_c)+3 m_c^6 \zeta _{1,-2}(m_c)-6 m_c^4 M^2 \zeta
_{1,-2}(m_c)+6 m_c^4 M^2 \zeta _{1,-1}(m_c)\nonumber\\
&-&12 m_c^2 M^4 \zeta _{1,-1}(m_c)+6 m_c^2 M^4
\zeta _{1,0}(m_c)-12 M^6 \zeta _{1,0}(m_c)+m_c^6 \zeta _{2,-3}(m_c)+3 m_c^4 M^2 \zeta _{2,-2}(m_c)\nonumber\\
&+&6
m_c^2 M^4 \zeta _{2,-1}(m_c)+6 M^6 \zeta _{2,0}(m_c)\Bigg)+2 m_\rho^2 \Bigg(-8 m_c^4 q^2
u_0 i_1({\cal T}_4(\alpha_i),1) \zeta _{-2,0}(m_c)\nonumber\\
&-&7 m_c^4 M^2 i_2({\cal S}(\alpha_i),1) \zeta _{-2,0}(m_c)+8
m_c^4 M^2 i_2({\cal S}(\alpha_i),v) \zeta _{-2,0}(m_c)\nonumber\\
&+&3 m_c^4 M^2 i_2(\tilde{S}(\alpha_i),1) \zeta _{-2,0}(m_c)+m_c^4
M^2 i_2({\cal T}_1(\alpha_i),1) \zeta _{-2,0}(m_c)\nonumber\\
&-&2 m_c^4 M^2 i_2({\cal T}_1(\alpha_i),v) \zeta _{-2,0}(m_c)-m_c^4
M^2 i_2({\cal T}_2(\alpha_i),1) \zeta _{-2,0}(m_c)\nonumber\\
&+&2 m_c^4 M^2 i_2({\cal T}_2(\alpha_i),v) \zeta _{-2,0}(m_c)-2
m_c^4 M^2 i_2({\cal T}_3(\alpha_i),v) \zeta _{-2,0}(m_c)\nonumber\\
&-&m_c^4 M^2 i_2({\cal T}_4(\alpha_i),1) \zeta _{-2,0}(m_c)-m_c^2 M^4 i_2({\cal T}_1(\alpha_i),1) \zeta _{0,-1}(m_c)\nonumber\\
&+&2
m_c^4 M^2 i_2({\cal T}_4(\alpha_i),v) \zeta _{-2,0}(m_c)-10 m_c^4 q^2 u_0 i_1({\cal T}_4(\alpha_i),1)
\zeta _{-1,-1}(m_c)\nonumber\\
&-&13 m_c^4 M^2 i_2({\cal S}(\alpha_i),1) \zeta _{-1,-1}(m_c)+16 m_c^4 M^2 i_2({\cal S}(\alpha_i),v)
\zeta _{-1,-1}(m_c)\nonumber\\
&+&3 m_c^4 M^2 i_2(\tilde{S}(\alpha_i),1) \zeta _{-1,-1}(m_c)+2 m_c^4 M^2 i_2({\cal T}_1(\alpha_i),1)
\zeta _{-1,-1}(m_c)\nonumber\\
&-&4 m_c^4 M^2 i_2({\cal T}_1(\alpha_i),v) \zeta _{-1,-1}(m_c)-2 m_c^4 M^2 i_2({\cal T}_2(\alpha_i),1)
\zeta _{-1,-1}(m_c)\nonumber\\
&+&4 m_c^4 M^2 i_2({\cal T}_2(\alpha_i),v) \zeta _{-1,-1}(m_c)+m_c^4 M^2 i_2({\cal T}_3(\alpha_i),1)
\zeta _{-1,-1}(m_c)\nonumber\\
&-&4 m_c^4 M^2 i_2({\cal T}_3(\alpha_i),v) \zeta _{-1,-1}(m_c)-m_c^4 M^2 i_2({\cal T}_4(\alpha_i),1)
\zeta _{-1,-1}(m_c)\nonumber\\
&+&4 m_c^4 M^2 i_2({\cal T}_4(\alpha_i),v) \zeta _{-1,-1}(m_c)+4 m_c^4 q^2 u_0
i_1({\cal T}(\alpha_i),1) (\zeta _{-2,0}(m_c)+\zeta _{-1,-1}(m_c))\nonumber\\
&-&6 m_c^4 q^2 u_0 i_1({\cal T}_3(\alpha_i),1)
(\zeta _{-2,0}(m_c)+\zeta _{-1,-1}(m_c))-32 m_c^4 M^2 u_0 \psi_3^\parallel(u_0) \zeta _{-1,0}(m_c)\nonumber\\
&+&48
m_c^2 M^4 u_0 \psi_3^\parallel(u_0) \zeta _{-1,0}(m_c)+2 m_c^4 q^2 u_0 i_1({\cal T}_4(\alpha_i),1)
\zeta _{-1,0}(m_c)\nonumber\\
&+&2 m_c^2 M^2 q^2 u_0 i_1({\cal T}_4(\alpha_i),1) \zeta _{-1,0}(m_c)+6 m_c^4 M^2
i_2({\cal S}(\alpha_i),1) \zeta _{-1,0}(m_c)\nonumber\\
&+&6 m_c^2 M^4 i_2({\cal S}(\alpha_i),1) \zeta _{-1,0}(m_c)-8 m_c^4
M^2 i_2({\cal S}(\alpha_i),v) \zeta _{-1,0}(m_c)\nonumber\\
&-&8 m_c^2 M^4 i_2({\cal S}(\alpha_i),v) \zeta _{-1,0}(m_c)-m_c^4
M^2 i_2({\cal T}_1(\alpha_i),1) \zeta _{-1,0}(m_c)\nonumber\\
&-&m_c^2 M^4 i_2({\cal T}_1(\alpha_i),1) \zeta _{-1,0}(m_c)+2
m_c^4 M^2 i_2({\cal T}_1(\alpha_i),v) \zeta _{-1,0}(m_c)\nonumber\\
&+&2 m_c^2 M^4 i_2({\cal T}_1(\alpha_i),v) \zeta _{-1,0}(m_c)+m_c^4
M^2 i_2({\cal T}_2(\alpha_i),1) \zeta _{-1,0}(m_c)\nonumber\\
&+&m_c^2 M^4 i_2({\cal T}_2(\alpha_i),1) \zeta _{-1,0}(m_c)-2
m_c^4 M^2 i_2({\cal T}_2(\alpha_i),v) \zeta _{-1,0}(m_c)\nonumber\\
&-&2 m_c^2 M^4 i_2({\cal T}_2(\alpha_i),v) \zeta _{-1,0}(m_c)-m_c^4
M^2 i_2({\cal T}_3(\alpha_i),1) \zeta _{-1,0}(m_c)\nonumber\\
&-&m_c^2 M^4 i_2({\cal T}_3(\alpha_i),1) \zeta _{-1,0}(m_c)+2
m_c^4 M^2 i_2({\cal T}_3(\alpha_i),v) \zeta _{-1,0}(m_c)\nonumber\\
&+&2 m_c^2 M^4 i_2({\cal T}_3(\alpha_i),v) \zeta _{-1,0}(m_c)-2
m_c^4 M^2 i_2({\cal T}_4(\alpha_i),v) \zeta _{-1,0}(m_c)\nonumber\\
&-&2 m_c^2 M^4 i_2({\cal T}_4(\alpha_i),v) \zeta _{-1,0}(m_c)-2
m_c^4 q^2 u_0 i_1({\cal T}_4(\alpha_i),1) \zeta _{0,-2}(m_c)\nonumber\\
&-&6 m_c^4 M^2 i_2({\cal S}(\alpha_i),1)
\zeta _{0,-2}(m_c)+8 m_c^4 M^2 i_2({\cal S}(\alpha_i),v) \zeta _{0,-2}(m_c)\nonumber\\
&+&m_c^4 M^2 i_2({\cal T}_1(\alpha_i),1)
\zeta _{0,-2}(m_c)-2 m_c^4 M^2 i_2({\cal T}_1(\alpha_i),v) \zeta _{0,-2}(m_c)\nonumber\\
&-&m_c^4 M^2 i_2({\cal T}_2(\alpha_i),1)
\zeta _{0,-2}(m_c)+2 m_c^4 M^2 i_2({\cal T}_2(\alpha_i),v) \zeta _{0,-2}(m_c)\nonumber\\
&+&m_c^4 M^2 i_2({\cal T}_3(\alpha_i),1)
\zeta _{0,-2}(m_c)-2 m_c^4 M^2 i_2({\cal T}_3(\alpha_i),v) \zeta _{0,-2}(m_c)\nonumber\\
&+&2 m_c^4 M^2 i_2({\cal T}_4(\alpha_i),v)
\zeta _{0,-2}(m_c)-48 m_c^4 M^2 u_0 \psi_3^\parallel(u_0) \zeta _{0,-1}(m_c)\nonumber\\
&+&48 m_c^2 M^4 u_0
\psi_3^\parallel(u_0) \zeta _{0,-1}(m_c)+4 m_c^4 q^2 u_0 i_1({\cal T}_4(\alpha_i),1) \zeta _{0,-1}(m_c)\nonumber\\
&+&2
m_c^2 M^2 q^2 u_0 i_1({\cal T}_4(\alpha_i),1) \zeta _{0,-1}(m_c)+12 m_c^4 M^2 i_2({\cal S}(\alpha_i),1)
\zeta _{0,-1}(m_c)\nonumber\\
&+&6 m_c^2 M^4 i_2({\cal S}(\alpha_i),1) \zeta _{0,-1}(m_c)-16 m_c^4 M^2 i_2({\cal S}(\alpha_i),v)
\zeta _{0,-1}(m_c)\nonumber\\
&-&8 m_c^2 M^4 i_2({\cal S}(\alpha_i),v) \zeta _{0,-1}(m_c)-2 m_c^4 M^2 i_2({\cal T}_1(\alpha_i),1)
\zeta _{0,-1}(m_c)\nonumber\\
&+&4 m_c^4 M^2 i_2({\cal T}_1(\alpha_i),v)
\zeta _{0,-1}(m_c)+2 m_c^2 M^4 i_2({\cal T}_1(\alpha_i),v) \zeta _{0,-1}(m_c)\nonumber\\
&+&2 m_c^4 M^2 i_2({\cal T}_2(\alpha_i),1)
\zeta _{0,-1}(m_c)+m_c^2 M^4 i_2({\cal T}_2(\alpha_i),1) \zeta _{0,-1}(m_c)\nonumber\\
&-&4 m_c^4 M^2 i_2({\cal T}_2(\alpha_i),v)
\zeta _{0,-1}(m_c)-2 m_c^2 M^4 i_2({\cal T}_2(\alpha_i),v) \zeta _{0,-1}(m_c)\nonumber\\
&-&2 m_c^4 M^2 i_2({\cal T}_3(\alpha_i),1)
\zeta _{0,-1}(m_c)-m_c^2 M^4 i_2({\cal T}_3(\alpha_i),1) \zeta _{0,-1}(m_c)\nonumber\\
&+&4 m_c^4 M^2 i_2({\cal T}_3(\alpha_i),v)
\zeta _{0,-1}(m_c)+2 m_c^2 M^4 i_2({\cal T}_3(\alpha_i),v) \zeta _{0,-1}(m_c)\nonumber\\
&-&4 m_c^4 M^2 i_2({\cal T}_4(\alpha_i),v)
\zeta _{0,-1}(m_c)-2 m_c^2 M^4 i_2({\cal T}_4(\alpha_i),v) \zeta _{0,-1}(m_c)+16 m_c^4 M^2 u_0
\psi_3^\parallel(u_0) \zeta _{0,0}(m_c)\nonumber\\
&-&128 m_c^2 M^4 u_0 \psi_3^\parallel(u_0) \zeta _{0,0}(m_c)+96
M^6 u_0 \psi_3^\parallel(u_0) \zeta _{0,0}(m_c)+4 m_c^2 M^2 q^2 u_0 i_1({\cal T}_4(\alpha_i),1)
\zeta _{0,0}(m_c)\nonumber\\
&+&2 M^4 q^2 u_0 i_1({\cal T}_4(\alpha_i),1) \zeta _{0,0}(m_c)+12 m_c^2 M^4
i_2({\cal S}(\alpha_i),1) \zeta _{0,0}(m_c)+6 M^6 i_2({\cal S}(\alpha_i),1) \zeta _{0,0}(m_c)\nonumber\\
&-&16 m_c^2 M^4
i_2({\cal S}(\alpha_i),v) \zeta _{0,0}(m_c)-8 M^6 i_2({\cal S}(\alpha_i),v) \zeta _{0,0}(m_c)-2 m_c^2 M^4
i_2({\cal T}_1(\alpha_i),1) \zeta _{0,0}(m_c)\nonumber\\
&-&M^6 i_2({\cal T}_1(\alpha_i),1) \zeta _{0,0}(m_c)+4 m_c^2 M^4
i_2({\cal T}_1(\alpha_i),v) \zeta _{0,0}(m_c)+2 M^6 i_2({\cal T}_1(\alpha_i),v) \zeta _{0,0}(m_c)\nonumber\\
&+&2 m_c^2 M^4
i_2({\cal T}_2(\alpha_i),1) \zeta _{0,0}(m_c)+M^6 i_2({\cal T}_2(\alpha_i),1) \zeta _{0,0}(m_c)-4 m_c^2 M^4
i_2({\cal T}_2(\alpha_i),v) \zeta _{0,0}(m_c)\nonumber\\
&-&2 M^6 i_2({\cal T}_2(\alpha_i),v) \zeta _{0,0}(m_c)-2 m_c^2 M^4
i_2({\cal T}_3(\alpha_i),1) \zeta _{0,0}(m_c)-M^6 i_2({\cal T}_3(\alpha_i),1) \zeta _{0,0}(m_c)\nonumber\\
&+&4 m_c^2 M^4
i_2({\cal T}_3(\alpha_i),v) \zeta _{0,0}(m_c)+2 M^6 i_2({\cal T}_3(\alpha_i),v) \zeta _{0,0}(m_c)-4 m_c^2 M^4
i_2({\cal T}_4(\alpha_i),v) \zeta _{0,0}(m_c)\nonumber\\
&-&2 M^6 i_2({\cal T}_4(\alpha_i),v) \zeta _{0,0}(m_c)-16 m_c^4 M^2
u_0 \psi_3^\parallel(u_0) \zeta _{1,-2}(m_c)+2 m_c^4 q^2 u_0 i_1({\cal T}_4(\alpha_i),1) \zeta _{1,-2}(m_c)\nonumber\\
&+&6
m_c^4 M^2 i_2({\cal S}(\alpha_i),1) \zeta _{1,-2}(m_c)-8 m_c^4 M^2 i_2({\cal S}(\alpha_i),v) \zeta _{1,-2}(m_c)\nonumber\\
&-&m_c^4
M^2 i_2({\cal T}_1(\alpha_i),1) \zeta _{1,-2}(m_c)+2 m_c^4 M^2 i_2({\cal T}_1(\alpha_i),v) \zeta _{1,-2}(m_c)\nonumber\\
&+&m_c^4
M^2 i_2({\cal T}_2(\alpha_i),1) \zeta _{1,-2}(m_c)-2 m_c^4 M^2 i_2({\cal T}_2(\alpha_i),v) \zeta _{1,-2}(m_c)\nonumber\\
&-&m_c^4
M^2 i_2({\cal T}_3(\alpha_i),1) \zeta _{1,-2}(m_c)+2 m_c^4 M^2 i_2({\cal T}_3(\alpha_i),v) \zeta _{1,-2}(m_c)\nonumber\\
&-&2
m_c^4 M^2 i_2({\cal T}_4(\alpha_i),v) \zeta _{1,-2}(m_c)+32 m_c^4 M^2 u_0 \psi_3^\parallel(u_0) \zeta
_{1,-1}(m_c)\nonumber\\
&-&112 m_c^2 M^4 u_0 \psi_3^\parallel(u_0) \zeta _{1,-1}(m_c)+4 m_c^2 M^2 q^2
u_0 i_1({\cal T}_4(\alpha_i),1) \zeta _{1,-1}(m_c)\nonumber\\
&+&12 m_c^2 M^4 i_2({\cal S}(\alpha_i),1) \zeta _{1,-1}(m_c)-16
m_c^2 M^4 i_2({\cal S}(\alpha_i),v) \zeta _{1,-1}(m_c)\nonumber\\
&-&2 m_c^2 M^4 i_2({\cal T}_1(\alpha_i),1) \zeta
_{1,-1}(m_c)+4 m_c^2 M^4 i_2({\cal T}_1(\alpha_i),v) \zeta _{1,-1}(m_c)\nonumber\\
&+&2 m_c^2 M^4 i_2({\cal T}_2(\alpha_i),1)
\zeta _{1,-1}(m_c)-4 m_c^2 M^4 i_2({\cal T}_2(\alpha_i),v) \zeta _{1,-1}(m_c)\nonumber\\
&-&2 m_c^2 M^4 i_2({\cal T}_3(\alpha_i),1)
\zeta _{1,-1}(m_c)+4 m_c^2 M^4 i_2({\cal T}_3(\alpha_i),v) \zeta _{1,-1}(m_c)\nonumber\\
&-&4 m_c^2 M^4 i_2({\cal T}_4(\alpha_i),v)
\zeta _{1,-1}(m_c)+64 m_c^2 M^4 u_0 \psi_3^\parallel(u_0) \zeta _{1,0}(m_c)\nonumber\\
&-&192 M^6 u_0 \psi_3^\parallel(u_0)
\zeta _{1,0}(m_c)+4 M^4 q^2 u_0 i_1({\cal T}_4(\alpha_i),1) \zeta _{1,0}(m_c)\nonumber\\
&+&12 M^6 i_2({\cal S}(\alpha_i),1)
\zeta _{1,0}(m_c)-16 M^6 i_2({\cal S}(\alpha_i),v) \zeta _{1,0}(m_c)-2 M^6 i_2({\cal T}_1(\alpha_i),1) \zeta
_{1,0}(m_c)\nonumber\\
&+&4 M^6 i_2({\cal T}_1(\alpha_i),v) \zeta _{1,0}(m_c)+2 M^6 i_2({\cal T}_2(\alpha_i),1) \zeta _{1,0}(m_c)-4
M^6 i_2({\cal T}_2(\alpha_i),v) \zeta _{1,0}(m_c)\nonumber\\
&-&2 M^6 i_2({\cal T}_3(\alpha_i),1) \zeta _{1,0}(m_c)+4 M^6
i_2({\cal T}_3(\alpha_i),v) \zeta _{1,0}(m_c)-4 M^6 i_2({\cal T}_4(\alpha_i),v) \zeta _{1,0}(m_c)\nonumber\\
&+&16 m_c^4 M^2
u_0 \psi_3^\parallel(u_0) \zeta _{2,-2}(m_c)+64 m_c^2 M^4 u_0 \psi_3^\parallel(u_0) \zeta _{2,-1}(m_c)+96
M^6 u_0 \psi_3^\parallel(u_0) \zeta _{2,0}(m_c)\Bigg)\Bigg]\Biggr\}~,\nonumber\\
\end{eqnarray}

\begin{eqnarray}
\Pi_B^{0}&=&\frac{ e^{-\frac{q^2}{M_1^2+M_2^2}}}{24
	\pi ^2} \Bigg\{4 m_c f_\rho^\parallel m_\rho \bigg[4 m_\rho^2\big (i_1(\phi(\alpha_i),1)+i_1(\psi(\alpha_i),1)\big)
\big(\zeta _{-1,0}(m_c)-\zeta _{0,0}(m_c) \big)\nonumber\\
&+&M^2 \psi_3^\perp(u_0) \big(-\zeta _{0,0}(m_c)+\zeta _{1,0}(m_c)\big)\bigg]+f_\rho^\perp
\bigg[-2 M^2 m_\rho^2 \bigg(i_2({\cal S}(\alpha_i),1)+3 i_2(\tilde{{\cal S}}(\alpha_i),1)\nonumber\\
&-&4 i_2(\tilde{{\cal S}}(\alpha_i),v)+i_2({\cal T}_1(\alpha_i),1)-2
i_2({\cal T}_1(\alpha_i),v)-i_2({\cal T}_2(\alpha_i),1)+2 i_2({\cal T}_2(\alpha_i),v)\nonumber\\
&+&2 i_2({\cal T}_3(\alpha_i),1)-4 i_2({\cal T}_3(\alpha_i),v)-i_2({\cal T}_4(\alpha_i),1)+2
i_2({\cal T}_4(\alpha_i),v)\bigg) \big[\zeta _{0,0}(m_c)-\zeta _{1,0}(m_c)\big]\nonumber\\
&+&4 M^4 \phi^\perp(u_0) (-\zeta
_{1,0}(m_c)+\zeta _{2,0}(m_c))+{\cal A}^\perp(u_0) m_\rho^2 (m_c^2 \zeta _{0,0}(m_c)+m_c^2
\zeta _{1,-1}(m_c)\nonumber\\
&-&m_c^2 \zeta _{1,0}(m_c)+3 M^2 \zeta _{1,0}(m_c)-m_c^2 \zeta _{2,-1}(m_c)-3 M^2
\zeta _{2,0}(m_c))+16 \tilde{j}_1({\cal B}^\perp(u)) m_\rho^2 \bigg(3 m_c^2 \zeta _{0,0}(m_c)\nonumber\\
&+&2 m_c^2 \zeta _{1,-1}(m_c)-2
\big[(m_c^2-4 M^2) \zeta _{1,0}(m_c)+m_c^2 \zeta _{2,-1}(m_c)+4 M^2 \zeta _{2,0}(m_c)\big]\bigg)\bigg]\Bigg\}~,
\end{eqnarray}

\begin{eqnarray}
\Pi_B^{GG}&=&-\frac{e^{-\frac{q^2}{M_1^2+M_2^2}}}{864 m_c M^6 \pi ^2} g_s^2\langle GG \rangle \Bigg\{m_c {\cal A}^\perp(u_0) f_\rho^\perp m_\rho^2
\Bigg[m_c^4 \zeta _{-2,0}(m_c)+2 m_c^4 \zeta _{-1,-1}(m_c)\nonumber\\
&-&m_c^4 \zeta _{-1,0}(m_c)+2 m_c^2 M^2
\zeta _{-1,0}(m_c)+m_c^4 \zeta _{0,-2}(m_c)-2 m_c^4 \zeta _{0,-1}(m_c)\nonumber\\
&+&2 m_c^2 M^2 \zeta _{0,-1}(m_c)-2
m_c^2 M^2 \zeta _{0,0}(m_c)+2 M^4 \zeta _{0,0}(m_c)-m_c^4 \zeta _{1,-2}(m_c)\nonumber\\
&-&2 m_c^2 M^2
\zeta_{1,-1}(m_c)-2 M^4 \zeta _{1,0}(m_c)\Bigg]+m_c f_\rho^\perp \Bigg[16 \tilde{j}_1({\cal B}^\perp(u)) m_\rho^2
\bigg(3 m_c^4 \zeta _{-2,0}(m_c)\nonumber\\
&+&6 m_c^4 \zeta _{-1,-1}(m_c)-3 m_c^4 \zeta _{-1,0}(m_c)+5 m_c^2
M^2 \zeta _{-1,0}(m_c)+3 m_c^4 \zeta _{0,-2}(m_c)\nonumber\\
&-&6 m_c^4 \zeta _{0,-1}(m_c)+5 m_c^2 M^2
\zeta _{0,-1}(m_c)-8 m_c^2 M^2 \zeta _{0,0}(m_c)+7 M^4 \zeta _{0,0}(m_c)\nonumber\\
&-&3 m_c^4 \zeta _{1,-2}(m_c)-8
m_c^2 M^2 \zeta _{1,-1}(m_c)-10 M^4\zeta _{1,0}(m_c)\bigg)-M^2 \bigg(m_\rho^2 (i_2({\cal S}(\alpha_i),1)\nonumber\\
&+&3i_2(\tilde{S}(\alpha_i),1)-4 i_2(\tilde{S}(\alpha_i),v)+i_2({\cal T}_1(\alpha_i),1)-2 i_2({\cal T}_1(\alpha_i),v)-i_2({\cal T}_2(\alpha_i),1)\nonumber\\
&+&2i_2({\cal T}_2(\alpha_i),v)+2 i_2({\cal T}_3(\alpha_i),1)-4 i_2({\cal T}_3(\alpha_i),v)-i_2({\cal T}_4(\alpha_i),1)\nonumber\\
&+&2 i_2({\cal T}_4(\alpha_i),v))
(m_c^2 (\zeta _{-1,0}(m_c)+\zeta _{0,-1}(m_c))+M^2 \zeta _{0,0}(m_c))\nonumber\\
&+&4 M^2 \phi^\perp(u_0)
(m_c^2 \zeta _{-1,0}(m_c)+m_c^2 \zeta _{0,-1}(m_c)-m_c^2 \zeta _{0,0}(m_c)+2 M^2 \zeta _{0,0}(m_c)\nonumber\\
&-&m_c^2
\zeta _{1,-1}(m_c)-2 M^2 \zeta _{1,0}(m_c))\bigg)\Bigg]+2 f_\rho^\parallel m_\rho \Bigg[4 m_c^4
m_\rho^2 (i_1(\phi(\alpha_i),1)\nonumber\\
&+&i_1(\psi(\alpha_i),1)) (\zeta _{-2,0}(m_c)+\zeta _{-1,-1}(m_c))+M^2
\psi_3^\perp(u_0) \bigg((-2 m_c^4+3 m_c^2 M^2) \zeta _{-1,0}(m_c)\nonumber\\
&+&3 m_c^2 (-m_c^2+M^2)
\zeta _{0,-1}(m_c)+m_c^4 \zeta _{0,0}(m_c)-8 m_c^2 M^2 \zeta _{0,0}(m_c)+6 M^4 \zeta _{0,0}(m_c)\nonumber\\
&-&m_c^4
\zeta _{1,-2}(m_c)+2 m_c^4 \zeta _{1,-1}(m_c)-7 m_c^2 M^2 \zeta _{1,-1}(m_c)+4 m_c^2 M^2
\zeta _{1,0}(m_c)\nonumber\\
&-&12 M^4 \zeta _{1,0}(m_c)+m_c^4 \zeta _{2,-2}(m_c)+4 m_c^2 M^2 \zeta _{2,-1}(m_c)+6
M^4 \zeta _{2,0}(m_c)\bigg)\Bigg]\Bigg\}~,
\end{eqnarray}
  
\begin{eqnarray}
\Pi_C^{0}&=&\frac{ e^{-\frac{q^2}{M_1^2+M_2^2}}}{3 \pi ^2} M^2 f_\rho^\parallel m_\rho \Bigg\{\Bigg[i_2(\alpha_g \tilde{{\cal A}}(\alpha_i),v)-2
i_2(\alpha_g \tilde{{\cal A}}(\alpha_i),v^2)+i_2(\alpha_q \tilde{{\cal A}}(\alpha_i),1)\nonumber\\
&-&2 i_2(\alpha_q \tilde{{\cal A}}(\alpha_i),v)-i_2(\alpha_g
{\cal V}(\alpha_i),v)-i_2(\alpha_q {\cal V}(\alpha_i),1)\Bigg] \bigg(\zeta _{0,0}(m_c)-\zeta _{1,0}(m_c)\bigg)\nonumber\\
&+&u_0^2
\psi_3^\perp(u_0) \bigg(\zeta _{1,0}(m_c)-\zeta _{2,0}(m_c)\bigg)\Bigg\}~,
\end{eqnarray}

\begin{eqnarray}
\Pi_C^{GG}&=&-\frac{ e^{-\frac{q^2}{M_1^2+M_2^2}}}{216
	M^4 \pi ^2} g_s^2\langle GG\rangle f_\rho^\parallel \Bigg\{\Bigg(i_2(\alpha_g \tilde{{\cal A}}(\alpha_i),v)-2
i_2(\alpha_g \tilde{{\cal A}}(\alpha_i),v^2)+i_2(\alpha_q \tilde{{\cal A}}(\alpha_i),1)\nonumber\\
&-&2 i_2(\alpha_q \tilde{{\cal A}}(\alpha_i),v)-i_2(\alpha_g
{\cal V}(\alpha_i),v)-i_2(\alpha_q {\cal V}(\alpha_i),1)\Bigg) \Bigg[m_c^2 (\zeta _{-1,0}(m_c)+\zeta _{0,-1}(m_c))\nonumber\\
&+&M^2
\zeta _{0,0}(m_c)\Bigg]+2 u_0^2 \psi_3^\perp(u_0) \Bigg[m_c^2 \zeta _{-1,0}(m_c)+m_c^2 \zeta _{0,-1}(m_c)-m_c^2
\zeta _{0,0}(m_c)+2 M^2 \zeta _{0,0}(m_c)\nonumber\\
&-&m_c^2 \zeta _{1,-1}(m_c)-2 M^2 \zeta _{1,0}(m_c)\Bigg]\Bigg\}~,
\end{eqnarray}

\begin{eqnarray}
\Pi_D^{0}&=&-\frac{2  e^{-\frac{q^2}{M_1^2+M_2^2}}}{3 M^2 \pi ^2} u_0 m_\rho^2 \Bigg\{4 m_c f_\rho^\parallel m_\rho \bigg(u_0
i_1(\psi(\alpha_i),1)-i_1(\alpha_g \psi(\alpha_i),v)\nonumber\\
&-&i_1(\alpha_q \psi(\alpha_i),1)\bigg) \big[\zeta _{-1,0}(m_c)-\zeta
_{0,0}(m_c)\big]+M^2 f_\rho^\perp \Bigg[\bigg(i_1({\cal T}(\alpha_i),1)\nonumber\\
&-&2 \big[i_1({\cal T}(\alpha_i),v)+i_1({\cal T}_4(\alpha_i),1)-2
i_1({\cal T}_4(\alpha_i),v)\big]\bigg) \bigg[\zeta _{0,0}(m_c)-\zeta _{1,0}(m_c)\bigg]\nonumber\\
&+&4 u_0 \tilde{j}_1({\cal B}^{\perp}(u))
\bigg(\zeta _{1,0}(m_c)-\zeta _{2,0}(m_c)\bigg)\Bigg]\Bigg\}~,
\end{eqnarray}
and
\begin{eqnarray}
\Pi_D^{GG} &=&\frac{ e^{-\frac{q^2}{M_1^2+M_2^2}}}{108
	M^8 \pi ^2} g_s^2\langle GG\rangle u_0 m_\rho^2 \Bigg\{4 m_c^3 f_\rho^\parallel m_\rho
\bigg[u_0 i_1(\psi,1)-i_1(\alpha_g \psi(\alpha_i),v)\nonumber\\
&-& i_1(\alpha_q \psi(\alpha_i),1)\bigg] \bigg[\zeta
_{-2,0}(m_c)+\zeta _{-1,-1}(m_c)\bigg]
+M^2 f_\rho^\perp\Bigg[\Bigg(i_1({\cal T}(\alpha_i),1)
\nonumber\\
&-&2 \bigg[i_1({\cal T}(\alpha_i),v)+i_1({\cal T}_4(\alpha_i),1)
-2i_1({\cal T}_4(\alpha_i),v)\bigg]\Bigg) \Bigg(m_c^2 \bigg[\zeta _{-1,0}(m_c)+\zeta _{0,-1}(m_c)\bigg]\nonumber\\
&+&M^2 \zeta
_{0,0}(m_c)\Bigg)+8 u_0 \tilde{j}_1({\cal B}^\perp (u))\Bigg(m_c^2 \zeta _{-1,0}(m_c)+m_c^2 \zeta _{0,-1}(m_c)-m_c^2
\zeta _{0,0}(m_c)\nonumber\\
&+&2 M^2 \zeta _{0,0}(m_c)-m_c^2 \zeta _{1,-1}(m_c)
-2 M^2 \zeta _{1,0}(m_c)\Bigg)\Bigg]\Bigg\}~.
\end{eqnarray}
In the above expressions the functions $\zeta_{m,n}(m_1,m_2)$ are defined as
\begin{eqnarray}
\zeta_{m,n}(m_1,m_2) = \int_{z_{\text{min}}}^{z_{\text{max}}}dz z^m \bar{z}^n e^{-\frac{m_1^2}{M^2 z}-\frac{m_2^2}{M^2 \bar{z}}}~,
\end{eqnarray}
where $m$ and $n$ are integers and $\zeta_{m,n}(m_1,m_1)=\zeta_{m,n}(m_1)$. We have also used the following short-hand notations:
\begin{eqnarray}
i_1(\phi(\alpha_i),f(v)) = \int {\cal D}\alpha_i \int_{0}^{1}dv \phi(\alpha_{q},\alpha_{\bar{q}},\alpha_{g}) f(v) \theta(k-u_0)~,
\end{eqnarray}
\begin{eqnarray}
i_2(\phi(\alpha_i),f(v)) = \int {\cal D}\alpha_i \int_{0}^{1}dv \phi(\alpha_{q},\alpha_{\bar{q}},\alpha_{g}) f(v) \delta(k-u_0)~,
\end{eqnarray}
\begin{eqnarray}
i_3(\phi(\alpha_i),f(v)) = \int {\cal D}\alpha_i \int_{0}^{1}dv \phi(\alpha_{q},\alpha_{\bar{q}},\alpha_{g}) f(v) \delta^{'}(k-u_0)~,
\end{eqnarray}
\begin{eqnarray}
j_1(f(u)) = \int_{u_0}^{1} du f(u)~,
\end{eqnarray}
and
\begin{eqnarray}
\tilde{j_1}(f(u)) = \int_{u_0}^{1} du (u-u_0) f(u)~,
\end{eqnarray}
where $k=\alpha_{q} - v\alpha_{g}$. For the explicit forms of wave functions of vector mesons and the functions used in the above expressions, in terms of  vector mesons' DAS, one can see Refs.  \cite{Ball:1998sk,Ball:1998ff,Ball:2007zt}.

\end{document}